\documentclass{aastex701}
\usepackage{mathtools}
\usepackage{xcolor}
\usepackage{graphicx}
\usepackage{amsmath}
\usepackage{physics}
\usepackage{mathptmx}
\usepackage{mathrsfs}
\usepackage{longtable, booktabs} 
\begin{document}

\title{Hydrodynamical simulations of helium-ignited binary white dwarf mergers}

\correspondingauthor{Vrutant Mehta, Robert Fisher}

\author{Vrutant Mehta}
\affiliation{Department of Physics, University of Massachusetts Dartmouth \\ 
285, Old Westport Road, North Dartmouth, MA, 02747, USA } 
\affiliation{Center for Scientific Computing and Data Science Research \\
University of Massachusetts, Dartmouth, MA 02747, USA } 
\email[show]{vmehta2@umassd.edu}

\author{Vishal Tiwari}
\affiliation{School of Physics, Georgia Institute of Technology \\
Howey Physics Bldg, 837 State St NW,  Atlanta, GA 30332, USA} 
\email{vtiwari@gatech.edu}

\author{Rüdiger Pakmor}
\affiliation{Max Planck Institute for Astrophysics \\
Karl-Schwarzschild-Strasse 1, Garching, 85748, Deutschland}
\email{rpakmor@mpa-garching.mpg.de}

\author{Divyanshu Singh}
\affiliation{Department of Physics, University of Massachusetts Dartmouth \\ 
285, Old Westport Road, North Dartmouth, MA, 02747, USA } 
\email{dsingh2@umassd.edu}

\author{Robert Fisher}
\affiliation{Department of Physics, University of Massachusetts Dartmouth \\
285, Old Westport Road, North Dartmouth, MA, 02747, USA }
\affiliation{Center for Scientific Computing and Data Science Research \\
University of Massachusetts, Dartmouth, MA 02747, USA } 
\email[show]{robert.fisher@umassd.edu}



\begin{abstract}

Type Ia supernovae (SNe Ia) are common luminous astrophysical transients. SNe Ia serve as distance indicators for measuring the expansion rate of the universe and play important roles in galactic nucleosynthesis. However, ambiguities persist regarding the nature of their stellar progenitors and explosion mechanisms. The recent discovery of \textit{Gaia} hypervelocity white dwarfs (WDs) has provided direct evidence in support of helium-ignited double degenerate SNe Ia. In this study, we investigate the outcomes of helium-ignited double-degenerate WD mergers by performing a set of 3D hydrodynamical simulations with two different codes: \texttt{AREPO} and \texttt{FLASH}. We consider two distinct binary WD systems close to helium ignition, evolving each with both codes while keeping initial conditions fixed. The first binary WD model produces a double detonation of the primary WD and the hypervelocity ejection of the surviving secondary, similar to the canonical dynamically driven double degenerate double detonation (D6) scenario. In the second model, the secondary also undergoes a core detonation, resulting in the complete disruption of both WDs. Notably, despite utilizing distinct numerical solvers, nuclear reaction networks, and mesh strategies, \texttt{AREPO} and \texttt{FLASH} produce broadly consistent outcomes for both sets of initial conditions. While the nucleosynthetic yields differ due to the different nuclear reaction networks employed, the overall agreement between the simulations demonstrates the robustness of the numerical modeling of this scenario. Our results strongly support the viability of both the D6 and quadruple detonation channels for at least some SNe Ia. We explore the prospective observational signatures of this channel, including in the X-rays using \textit{XRISM's} \textit {RESOLVE}.



\end{abstract}



\section{Introduction} \label{sec:introduction}
Type Ia supernovae (SNe Ia) are thought to result from the thermonuclear explosion of mass-accreting carbon-oxygen white dwarfs (WDs) in  binary systems. This inference is supported by observations of SN 2011fe and SN 2014J, which directly constrained the radius of the stellar progenitor to a compact object, and produced observable radioactive $^{56}$Co, respectively \citep{nugent2011supernova, bloom2011compact, chomiuketal12, churazov2014cobalt}. SNe Ia play an important role in galactic chemical evolution as sites of intermediate mass (IME) and iron group elements (IGEs) \citep{eitner2020observational}, and most famously as standardizable cosmological candles \citep{riess1998observational, schmidt_1998, perlmutter_1999}. Yet, despite the important role of SNe Ia in astrophysics and cosmology, the nature of their stellar progenitors and explosion mechanisms is still largely unknown.

Over the past few decades, the double-degenerate (DD) channel has received significant attention as a viable progenitor of SNe Ia \citep{santander2015double, jacobson2018constraining, moon2021rapidly, graham2022nebular, siebert2023asymmetric}. The DD channel, broadly conceived, consists of two sub-Chandrasekhar mass (sub-M$_{\rm Ch}$) WDs undergoing a dynamical merger. The DD channel accounts for the observed delay time distribution \citep{mennekensetal10, maozetal10}, the absence of luminous progenitors prior to explosion \citep{bloom2011compact, nugent2011supernova, brownetal12}, and the lack of dense circumstellar material in the post-explosion environment \citep{badenes2007models, horesh2012early} of SNe Ia. There are multiple routes through which the WD merger may ultimately lead to a SN Ia: a violent merger soon after complete or partial tidal disruption of the donor \citep{pakmor_2010, pakmor2026violent}, the collision of two white dwarfs \citep{rosswogetal09}, a spiral instability driven through the accretion disk of a completely disrupted donor \citep{kashyapetal15}, the detonation of helium atop the accretor through the dynamically driven double-degenerate double-detonation (D6) channel \citep{pakmor2013helium} that might also explode the donor immediately after in a similar way \citep{pakmor2022fate}, or even the core compression and ignition of the merger long after the inspiral \citep{raskin_2014}. 

Recent discoveries have provided direct evidence supporting the DD scenario in which the primary WD explodes well before the tidal disruption of the secondary WD. Notably, hypervelocity WDs $(v_{\mathrm{tangential}} \gtrsim 600 \ \mathrm{km \ s^{-1}} )$ identified in data from the \textit{Gaia} space telescope \citep{shenetal2018three, elbadryetal23} are consistent with being surviving companions ejected from binary systems following a SNe Ia. In particular, \cite{shenetal2018three} demonstrated that the trajectory of one such WD traces back to the known supernova remnant G70.0–21.5 within the Milky Way. Furthermore, some of these hypervelocity WDs exhibit unusual photospheric compositions--including trace amounts of heavy elements such as silicon, sulfur, nickel, and iron, in addition to carbon and oxygen \citep{chandra2022sn, werner2024photospheres, werner2024ultraviolet, hollands2025spectroscopic}--consistent with captured nuclear ash from a SN Ia. Collectively, these findings strongly suggest that such WDs are likely ex-companions from systems that underwent a D6 explosion.

In the D6 channel, a binary system of two intermediate mass stars undergoes two phases of common envelope evolution and leads to a tight binary of two sub-M$\rm_{Ch}$ mass ($0.5$ M$_{\odot}$ $ \lesssim$ M$_{\rm WD}$ $\lesssim 1.1$ M$_{\odot}$ ) carbon/oxygen WDs, each retaining a small amount ($\sim 0.005 - 0.01$ M$_{\odot}$) of unburned surface helium from main-sequence evolution. As the binary systems shrinks due to gravitational wave emission and two WDs approach Roche lobe overflow, the accretion stream from the secondary WD impacts the primary surface helium. Because helium is much more readily burned than carbon, the impacting accretion stream, with its kinetic energy dissipated as heat, eventually ignites and detonates the primary surface helium \citep{pakmor2013helium}. Once ignited, the helium runaway propagates through the surface helium, wrapping around the primary WD, and subsequently merging at the antipode of the ignition point. Simultaneously, the helium detonation drives compressive shocks into the WD interior. These inwardly propagating shocks transmit through the dense interior of the WD, with its higher sound speed, become refracted \citep{CourantFriedrichs48}, curving towards the center of the WD. Eventually, these compressive shocks converge at a single point near the core of the WD initiating a second detonation. Alternatively, the collision of helium detonation fronts at the antipode, can produce a strong shock moving in the direction of the WD center, and give rise to the second detonation via what is referred to as the ``scissors mechanism" by \cite{gronow2020sne}. However, this explosion differs from the shock convergence led explosion in regards to the location and timing for initiating the second detonation. If the core is detonated, it completely unbinds the primary WD, and synthesizes intermediate-mass and iron-group elements characteristic of a SN Ia. The secondary WD is necessarily very close to the primary in this channel, and is inevitably directly impacted by the detonation fronts \citep{piro2025early}, burning and ablating its outer layers, and partially depositing ejecta onto its surface. As originally proposed, the D6 channel predicts the secondary WD survives the explosion and is ejected roughly at its high orbital velocity at the time of ignition of the primary. This escaped companion can later be observed as a hypervelocity WD. These surviving hypervelocity WDs are important signatures of the D6 channel, but their scarcity in the \textit{Gaia} datasets implies that this mechanism by itself may account for only about $\sim 2\%$ of overall SNe Ia in the Milky Way Galaxy \citep{shen2025evolution}.

Thus far, only a few hydrodynamical simulations \citep{guillochon2010surface, pakmor2013helium, roy20223d, pakmor2022fate} have captured the a significant part or the full 3D end-to-end evolution of helium-ignited WD binaries ---from late inspiral and Roche-lobe overflow, through accretion-stream impact and helium-shell detonation, to shock convergence, possible core detonation, and homologous expansion of the ejecta. Most of those models \citep{guillochon2010surface, pakmor2013helium, roy20223d} followed the inspiral through accretion-stream impact and surface helium detonation, but did not achieve a core detonation in the primary WD. One possible explanation for this outcome lies in the initial conditions of the simulations. Earlier phases of those simulations including the inspiral and initiation of mass transfer were often carried out in smoothed-particle hydrodynamics (SPH) based codes, and then mapped onto a grid-based Eulerian codes (such as \texttt{FLASH} \citep{guillochon2010surface, roy20223d}, or a moving mesh code like \texttt{AREPO} \citep{pakmor2013helium}). Fundamentally, SPH codes represent fluids using mesh-free discrete particles in a Lagrangian framework. The SPH method follows the motion of the particles, and symmetrizes the pairwise gravitational force interactions between particles. In practice, the tree-methods \citep{barnes1986hierarchical} are often preferred to solve gravitational force between the particles, since it has shown excellent conservation of both linear and angular momentum \citep{springel2010smoothed}. However, because SPH is inherently a Lagrangian methodology, the low density of the helium envelope necessarily has large smoothing lengths, which will tend to lead to greater numerical viscosity, and an artificially large generation of entropy. Detonations initiated in these “puffed-up” helium layers propagate asymmetrically due to uneven distribution of the helium layer at lower speeds than in fully resolved layers. Such non-uniformly propagating detonation generates weaker shocks which adversely affect the conditions for core ignition. 

In contrast, \cite{pakmor2022fate} generated initial conditions from 1D profiles of both WDs, including their surface helium layers, mapped them into 3D with \texttt{AREPO} code \citep{springel2010pur, pakmor2016improving, weinberger2020arepo}, and then followed the full binary evolution. In this approach, they demonstrated that the ignition of a dense compact surface helium layer ($\sim$0.03 M$_{\odot}$) on the primary WD produces rapid, symmetric detonation and stronger inward shocks that lead to a successful core detonation along the locus of shock-shock convergence in the carbon-oxygen core. A recent study from \cite{pakmor2024type} has also emphasized that fully three-dimensional simulations are required to capture the hydrodynamic structures of the accretion stream, shock convergence, and the asymmetries in the ejecta from SNe Ia explosions. Reduced-dimensional models cannot represent the intrinsic 3D geometry of these features accurately, potentially failing to capture the critical detonation initiation in the surface layers or core, and leading to unrealistic ejecta structures and incorrect predictions of observable signatures.

Recent work has shown that most, if not all, carbon/oxygen WDs with masses $ \lesssim \rm 1.0 \ M_{\odot} $  have compositional profiles which can initiate surface helium detonations, and therefore host double detonations \citep{shen2024almost}. For more massive carbon/oxygen WDs ($\sim 1$ M$_{\odot}$), only $\sim 10^{-3}$ M$_{\odot}$ of mass accretion from the companion is sufficient to trigger the He-shell detonation, followed by the carbon/oxygen core detonation.
This also implies that the blast wave from the core detonation of the primary will also drive most secondaries to double detonate in helium-ignited WD binaries. Thus, helium-ignited binary WDs can potentially host not only double detonations SNe Ia, but also triple and quadruple detonations SNe Ia. Notably, quadruple detonation SNe Ia leave no surviving bound star, with relevance for the D6 Gaia rate problem.


The model from \cite{pakmor2022fate} has produced a core detonation of the primary WD, which in turn ignited a surface helium detonation on the secondary WD. Yet, the surface helium detonation on the secondary did not lead to the core detonation of the secondary WD. On the other hand, \cite{pakmor2021thermonuclear} simulated the 3D evolution of a carbon/oxygen WD accreting helium from a hybrid helium/carbon/oxygen companion. The accreted helium produced a surface detonation on the carbon/oxygen WD but failed to ignite its core. Instead, the helium detonation fronts burned through the accretion stream and initiated a surface helium detonation on the hybrid companion. The shocks produced by the helium detonation converge toward the center, successfully triggering a core detonation in the hybrid helium/carbon/oxygen companion.

These intriguing possible outcomes of helium-ignited binary WD mergers raise several key questions. Under what physical conditions does the primary WD double detonate? What happens to the secondary WD once the primary detonates? Does the secondary WD survive intact? If the secondary does survive, how is it enriched by the ejecta from the primary and burning in its envelope? If the secondary does not survive, under what conditions can it undergo helium and core detonation? How do triple and quadruple detonations affect the nucleosynthetic abundances in the ejecta? Finally, what distinctive spectroscopic and photometric features could differentiate the possible mechanisms of helium-ignited binary WD mergers from other Type Ia explosion scenarios?

To address these questions, we present in this paper 3D hydrodynamical simulations of helium-ignited WD mergers. To assess the robustness of the D6 mechanism, and to probe the reasons for the different outcomes of prior simulations, we undertake the simulations using two independent hydrodynamical codes: the \texttt{AREPO} moving-mesh code \citep{springel2010pur, pakmor2016improving, weinberger2020arepo} and the \texttt{FLASH} adaptive mesh refinement (AMR) Eulerian hydrodynamical code \citep{fryxell2000flash}. While both codes capture the same fundamental equations of Eulerian hydrodynamics, Poisson self-gravity, nuclear burning, and equation of state, they do so with fundamentally distinct discretization methodologies and implementations for hydrodynamics, self-gravity, nuclear burning. Any outcomes which are sensitive to numerical algorithms will be highlighted in the possible differences in outcomes of the two hydrodynamical frameworks. Likewise, agreement between the two codes will point towards results which are physically robust even in the presence of distinct numerical methodologies.

Our analysis emphasizes the comparison of outcomes between the two simulations, highlighting how helium shell ignition and core detonations proceed in the primary and secondary WDs under similar initial conditions. Consequently, the initial conditions for both sets of simulations are generated with the \texttt{AREPO} moving-mesh code \citep{weinberger2020arepo}. We note that the physical length scales at which the ignition and detonation of helium and carbon occurs are approximately four to five orders of magnitude smaller than the highest spatial resolutions in our simulations \citep{timmes2000cellular, holcomb2013conditions}. Thus, neither ignition-phases prior to the detonations nor the actual detonations are fully resolved in the simulations we present in this study.   

Section \ref{sec:methodology} outlines the detailed methodology employed in our simulations. The results are presented in Section \ref{sec:results}, where we also compare them with corresponding simulations performed using the \texttt{AREPO} code, including an analysis of the nucleosynthetic abundances obtained from both \texttt{AREPO} and \texttt{FLASH}. In Section \ref{sec:discussion}, we provide an in-depth examination of the current understanding of the helium-ignited binary WD merger channel and discuss the implications of our results for addressing key questions related to SNe Ia. Finally, Section \ref{sec:conclusion} summarizes the main findings of this study.

\section{Methodology} \label{sec:methodology}

In this paper, we present two different binary WD models to explore the fundamental questions regarding the outcomes of helium-ignited binary WD mergers. We carried out the simulations with each binary WD model independently with the \texttt{AREPO} \citep{springel2010pur, pakmor2016improving, weinberger2020arepo} and \texttt{FLASH} \citep{fryxell2000flash} hydrodynamical codes, allowing us to compare two self-contained outcomes for identical initial conditions. The first binary model configuration consists of a 1.0 M$_{\odot}$ primary WD and a 0.6 M$_{\odot}$ secondary WD, each with a 0.01 M$_{\odot}$ of helium layer, while the remaining mass in both WDs has a fixed carbon/oxygen composition of 50/50. The \texttt{AREPO} simulation with the first model resulted in a double detonation of the primary WD and ejection of the secondary WD with the hypervelocity (at Roche-lobe velocity). This outcome is consistent with the canonical D6 channel, as shown in \cite{pakmor2022fate}. We refer to this first binary model as the \texttt{Double\_Det} model hereafter. The second binary model configuration comprises a 1.0 M$_{\odot}$ primary WD with a 0.005 M$_{\odot}$ surface helium layer and a 0.6 M$_{\odot}$ secondary WD with a 0.03 M$_{\odot}$ helium layer on the surface, and the remaining mass in both WDs has a carbon/oxygen composition of 50/50. The simulation completed with the second model has produced quadruple detonations in \texttt{AREPO}. In this case, both WDs were completely disrupted, leaving no surviving companion. Throughout, we refer to this second model as the \texttt{Quad\_Det} model.

As we described in Section \ref{sec:introduction}, the mapping technique used to initialize mesh quantities in the simulation domain is critical. Unevenly distributed helium layers with high entropy can adversely affect shock speeds and, more importantly, the post-shock convergence density and temperature required to ignite the core detonation in the WD.
To ensure we accurately capture the surface helium layers on both WDs, the early phases of binary WD evolution are performed in \texttt{AREPO}, which resolves the surface helium layer with less artificial dissipation than prior SPH simulations \citep{pakmor2022fate}. We map the simulations to \texttt{FLASH} a few seconds before ignition in \texttt{AREPO} and follow it there as well. 



\subsection{{\rm AREPO} simulation setup}\label{subsec:AREPO_setup}

\texttt{AREPO} uses a second order finite volume scheme to solve the equations of hydrodynamics on a moving Voronoi mesh \citep{springel2010pur, pakmor2016improving,weinberger2020arepo}. The mesh-generating points that define the Voronoi mesh are moved with the local gas velocity and a correction term to keep the mesh regular. In this manner, the fluxes over interfaces are kept minimal, and the numerical diffusivity of the scheme is small.

The white dwarfs are initially set up independently in \texttt{AREPO} and actively relaxed for five dynamical timescales to ensure that they are and remain in hydrostatic equilibrium \citep{pakmor2012stellar,ohlmann2017constructing}. We use cells of roughly equal mass initially, and employ explicit refinement and de-refinement to make sure that the mass of cells stays within a factor of $2$ around the target gas mass of $10^{-7}\,\mathrm{M_\odot}$. We additionally refine cells that have a direct neighbor with a volume more than $10$ times smaller to avoid large resolution gradients in the grid. We use the Helmholtz equation of state \citep{timmes2000accuracy}. We include self-gravity via an oct-tree solver, and nuclear reactions via a nuclear reaction network that we solve on the fly and fully couple to the hydrodynamics \citep{pakmor2012stellar,pakmor2021thermonuclear}.

We then add both relaxed white dwarfs together in a binary system in co-rotation in a domain size of $10^{12}\,\mathrm{cm}$. We chose the initial separation large enough that the secondary white dwarf does not fill its Roche lobe. We then evolve the binary system with an artificial gravitational-wave like term that removes angular momentum from the system and shrinks the binary system at a constant rate of $100\,\mathrm{km/s}$ \citep{pakmor2021thermonuclear}. We continue this until the density at the inner Lagrange point of the binary system L1 reaches about $10^4\,\mathrm{g\,cm^{-3}}$. At that time, we deactivate the active inspiral term and continue to freely evolve the binary while conserving total angular momentum.

\subsection{{\rm FLASH} simulation setup}\label{subsec:FLASH_setup}

For our comparison simulations, we used the massively parallel \texttt{FLASH} hydrodynamical code \citep{fryxell2000flash} in full 3D geometry.
\texttt{FLASH} simulates the Eulerian equations of hydrodynamics, coupled to self-gravity and nuclear burning, and closed with the same Helmholtz free-energy equation of state (EOS) \citep{timmes2000accuracy} as for \texttt{AREPO}. We use \texttt{FLASH}'s unsplit higher-order Godunov scheme. The hydro updates utilize the piecewise parabolic method for the reconstruction, and compute fluxes at cell interfaces with the \texttt{HLLC} Riemann solver. We solve the Poisson equation for the gravitational potential, using an improved multipole expansion \citep{couch2013improved} with $l_{\rm max} = 60$. We set the boundary conditions for the simulation domain to \texttt{diode}, which ensures the zero gradient of fluxes at the boundary, while allowing mass and shocks to leave the simulation domain.  The nuclear reaction networks in \texttt{FLASH} were selected to match those used in \texttt{AREPO} as closely as possible. Specifically, the \texttt{Double\_Det} model, which was evolved with the $13-$isotope network \citep{pakmor2013helium, burmester2023arepo} in \texttt{AREPO}, is simulated with the closely-related \texttt{approx13} network \citep{timmes1999integration} in \texttt{FLASH}. These two networks both retain 13 isotopes along the $\alpha$-chain, though \texttt{approx13} retains $(\alpha,p)(p,\gamma) \ \text{and} \ (\gamma,p)(p,\alpha)$ proton-catalyzed side chains modeled in fast equilibrium with the $\alpha$-chain. For the \texttt{Quad\_Det} model, which used the more extensive $55-$isotope network \citep{pakmor2012stellar, pakmor2021thermonuclear, pakmor2022fate} in \texttt{AREPO}, we adopt the \texttt{approx19} network \citep{timmes1999integration, kashyapetal17, roy20223d} in \texttt{FLASH}. While
\texttt{approx19} is smaller than the $55-$isotope network used in \texttt{AREPO}, it incorporates the same proton-catalyzed side chains from \texttt{approx13}. In adapting the $55-$isotope \texttt{AREPO} abundances to the \texttt{approx19} network in \texttt{FLASH}, we included only species with mass fractions greater than $10^{-16}$, while preserving key $\alpha$-chain isotopes ($\mathrm{^{4}He, ^{12}C, ^{16}O, ^{20}Ne, ^{24}Mg, ^{28}Si}$) from the initial \texttt{AREPO} composition.


The \texttt{FLASH} simulations employ AMR refinement criteria designed to resolve the binary stars as well as the regions which give rise to detonation initiations, based on temperature and density thresholds. Specifically, in order to resolve each of the stars, blocks located within a spherical volume of $6000$ km radius around the WD centers (determined using cell with highest density on the computational domain) are refined by a factor of two whenever any cell in the block exceeds a density of $10^{5}$ g cm$^{-3}$. The center of this spherical volume is determined by locating the cell with the highest density on the Eulerian mesh of the computational domain. In addition to this, for refinement of regions with active burning a separate temperature-based refinement is applied to any block that contains at least one cell with a temperature above the threshold of $10^{9}$ K. Both criteria allows refinement up to one level coarser than our maximum refinement. The highest spatial resolution, with a linear resolution of $\sim 13.4$ km per cell, is enabled only for those blocks where the shock-shock convergence occurs in the vicinity of the center of the WD. Upon convergence, the density shows a sharp spike, and this additional refinement criterion is enabled. To implement this, we supplement our density-based refinement criterion with a second threshold--set to $\mathrm{5 \times 10^{7} \ g \ cm^{-3}}$ for the primary and $\mathrm{6.4 \times 10^{6} \ g \ cm^{-3}}$ for the secondary--slightly above the central densities of the corresponding WDs. As the shocks converge and compress the central region, the density of the affected cells surpasses this threshold, activating the highest refinement only in the small zone where the shock-shock convergence occurs. These criteria have effectively reduced the computational cost of simulation roughly by a factor of five, while still capturing model-predicted helium and carbon detonations.

In contrast to \texttt{AREPO}, which uses an unstructured Voronoi mesh, the \texttt{FLASH} simulation domain consists of a Cartesian oct-tree-based AMR grid with a factor of two refinement, and self-similar block structure. Therefore, mapping this unstructured mesh onto a structured grid in \texttt{FLASH} requires transformation of the cells due to their fundamentally different mesh sizes and shapes. To address the difference between mesh geometries, the \texttt{AREPO} mesh data are extracted from the unstructured Voronoi mesh and mapped onto a coarsened up uniform grid of $\rm 512^{3}$ cells, corresponding to a resolution of $\sim 107$ km. The domain spans from $-27,500$ km to $27,500$ km in each dimension, sufficient to accommodate both stars and the accretion stream, as well as to track the ejecta for approximately a second after the explosion. For computational efficiency, this dataset was carefully down-sampled first to $\rm 256^{3}$ cells and then further to $\rm 128^{3}$ cells, while preserving key quantities such as mass, momentum, energy, and abundances. The overall errors introduced by this down-sampling remain below $10^{-12}$ \% relative for mass, momentum, abundances, and $\lesssim 2\%$ for energy-related quantities (kinetic, thermal, gravitational, total). Importantly, this reduction in resolution preserves the local thermodynamic and compositional structures essential for resolving the fine-scale conditions leading to helium ignition. We verified this by following the  spatial location and time of the helium ignition in both the $\rm 512^{3}$ and down-sampled $\rm 128^{3}$ dataset, which remains consistent in both. This down-sampled $\rm 128^{3}$ data corresponds to the base level resolution of $\sim 430$ km on the same domain size.

By construction, the \texttt{FLASH} framework employs an Eulerian grid, where the hydrodynamic quantities are evolved within spatial cells rather than individual fluid elements, similar to a co-moving mesh code \texttt{AREPO}. To enable the tracking of localized fluid properties during the evolution, we introduce $10^{6}$ passive Lagrangian tracer particles distributed uniformly by mass throughout the computational domain. These tracers are passively advected with the flow, and record the local kinematic, thermodynamic, and compositional histories of the fluid, providing data for subsequent post-processing analyses.

The initial conditions for the \texttt{FLASH} simulations were extracted from an \texttt{AREPO} snapshot taken a few seconds before the onset of helium ignition on the primary surface, ensuring that the subsequent evolution is individual in \texttt{FLASH}, while the comparisons can also be made for the similar ignition and detonation states between \texttt{AREPO} and \texttt{FLASH}.

\section{Results} \label{sec:results}
To investigate the robustness of the helium-ignited binary white dwarf merger channel, we evolved each initial conditions for the binary systems using two independent hydrodynamical simulation codes. The resulting \texttt{FLASH} simulations are compared to the corresponding \texttt{AREPO} simulations with respect to their detonation outcomes, nucleosynthetic yields, and overall agreement in morphology, timing, and energetics at different stages of their evolution. Furthermore, we used the \texttt{FLASH} simulations to examine the underlying nuclear reactions which produced the helium layer ignition and detonation. We also investigated the interactions of resulting shocks, and the physical conditions produced by the convergence of shocks which led to the core detonation in our simulations. Previous studies have explored the idealized conditions for some of these analyses \citep{fink2007double, fink2010double, moll2013multi, shen2014ignition}. In contrast, the simulations presented in this research provide a more robust understanding for those condition using complete three-dimensional models.

Subsection \ref{subsec:double_det} describes our analysis methods and the results obtained from the \texttt{Double\_Det} model evolved in \texttt{FLASH} code. The initial conditions of this model were taken at roughly four seconds prior to the onset of helium-layer ignition in the \texttt{AREPO} simulation. We successfully obtained both the helium and core detonations of the primary WD in our \texttt{FLASH} simulation, which are consistent with those seen in the \texttt{AREPO} simulation. The secondary WD nearby to the explosion gets impacted by the primary WD ejecta, and undergoes surface helium layer burning. Nonetheless, in both simulations of this system, the secondary WD did not undergo core detonation and remained intact. We discuss the prospective implications of these surviving WDs in the context of \textit{Gaia}-hypervelocity WDs, in the Section (\ref{sec:discussion}). Figures \ref{fig:double_det_densities_evolution} and \ref{fig:double_det_temp_evolution} display the mid-plane slice-plots along $z-$axis showing the mass density, $^{4}\mathrm{He}-$density, temperature, and temperature gradient magnitude fields of the \texttt{Double\_Det} model. These plots compare the identical phases of the \texttt{AREPO} and \texttt{FLASH} simulations at multiple stages of evolution. 

Our \texttt{FLASH} simulation is limited by a smaller computational domain size in comparison to the \texttt{AREPO} simulation. Consequently, about a second after the core detonation of the primary WD, the resulting ejecta start to leave the computational domain. This limitation makes it difficult to quantify and predict the observable signatures from the \texttt{FLASH} simulations. Thus we do not compute any synthetic observable analysis for this model. However, we quantify and compare the abundances produced in the \texttt{AREPO} and \texttt{FLASH} simulations at approximately one second after the core detonation of primary WD. At this state, most of the initial mass ($\gtrsim 99 \%$) is still on the simulation domain. These abundances are reported in the Table \ref{tab:double_det_ejecta_comparison}.  


Our second model, referred to as the \texttt{Quad\_Det} model, differs from the \texttt{Double\_Det} model in both the initial helium layer masses of the binary WDs and the nuclear reaction networks which we used to evolve the system in both codes. 
The initial binary system consists of a $1\ \mathrm{M_{\odot}}$ primary and a $0.6\ \mathrm{M_{\odot}}$ secondary WD. These totals include a $0.005\ \mathrm{M_{\odot}}$ surface helium layer on the primary and a $0.03\ \mathrm{M_{\odot}}$ helium layer on the secondary WD, with the remaining core masses consisting of 50/50 carbon/oxygen. We use the \texttt{approx19} reaction network in \texttt{FLASH}, in contrast to the larger $55-$isotope network employed in \texttt{AREPO}, to capture the nuclear burning in the \texttt{Quad\_Det} model. The initial condition of this model for the \texttt{FLASH} simulation was taken approximately five seconds before the initialization of helium layer ignition on the primary WD in the \texttt{AREPO} simulation. 
In our \texttt{FLASH} simulation, we capture not only the double detonation of the primary WD but also the helium and core detonation of the secondary WD, similar to the detonation outcome seen in the \texttt{AREPO} simulation. Consequently, we extend our analysis workflow to include both detonations of the secondary WD, investigating all four detonations occurring in this model. We describe our findings from the investigation of \texttt{Quad\_Det} model in Subsection \ref{subsec:quad_det}.

In the \texttt{FLASH} simulation of \texttt{Quad\_Det} model, the helium detonation of the primary WD occurred roughly three seconds earlier than in \texttt{AREPO} simulation, thus the subsequent evolution is followed by a phase offset between the two. Figures \ref{fig:quad_det_densities_evolution} and \ref{fig:quad_det_temp_evolution} shows the mid-plane slice-plots of same fields in \texttt{Quad\_Det} model as shown for \texttt{Double\_Det} model. These plots draw a comparison between the \texttt{AREPO} and \texttt{FLASH} simulations at similar states of their evolution, that is at the same time relative to helium ignition on the primary. The same limitation of a smaller computational domain size in \texttt{FLASH} simulations also applies to \texttt{Quad\_Det} model, and thus the overall final abundances are not quantifiable as the ejecta start to leave the domain after the explosion of primary WD. We compare the abundances at the identical evolutionary phase when the primary WD is exploded and more than $90 \%$ of total initial mass is still on the domain. These abundances are reported in the Table \ref{tab:quad_det_ejecta_comparison}.

\subsection{Double Det model} \label{subsec:double_det}

\subsubsection{Helium layer ignition of the primary white dwarf}

\begin{figure}[htb!]
    
    \gridline{\fig{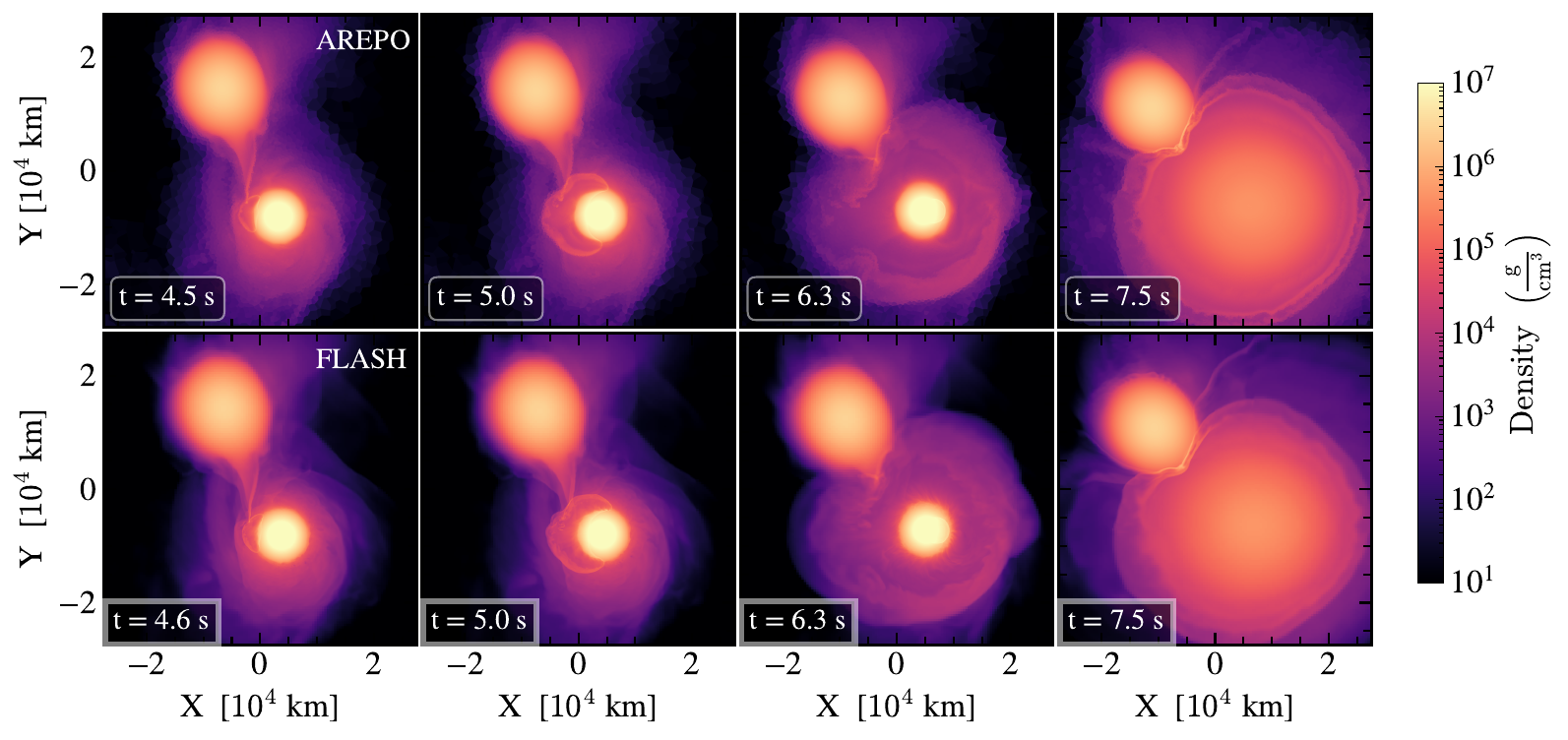}{0.95\linewidth}{(a) Mass density evolution in both AREPO (top row) and FLASH (bottom row).}}
    \gridline{\fig{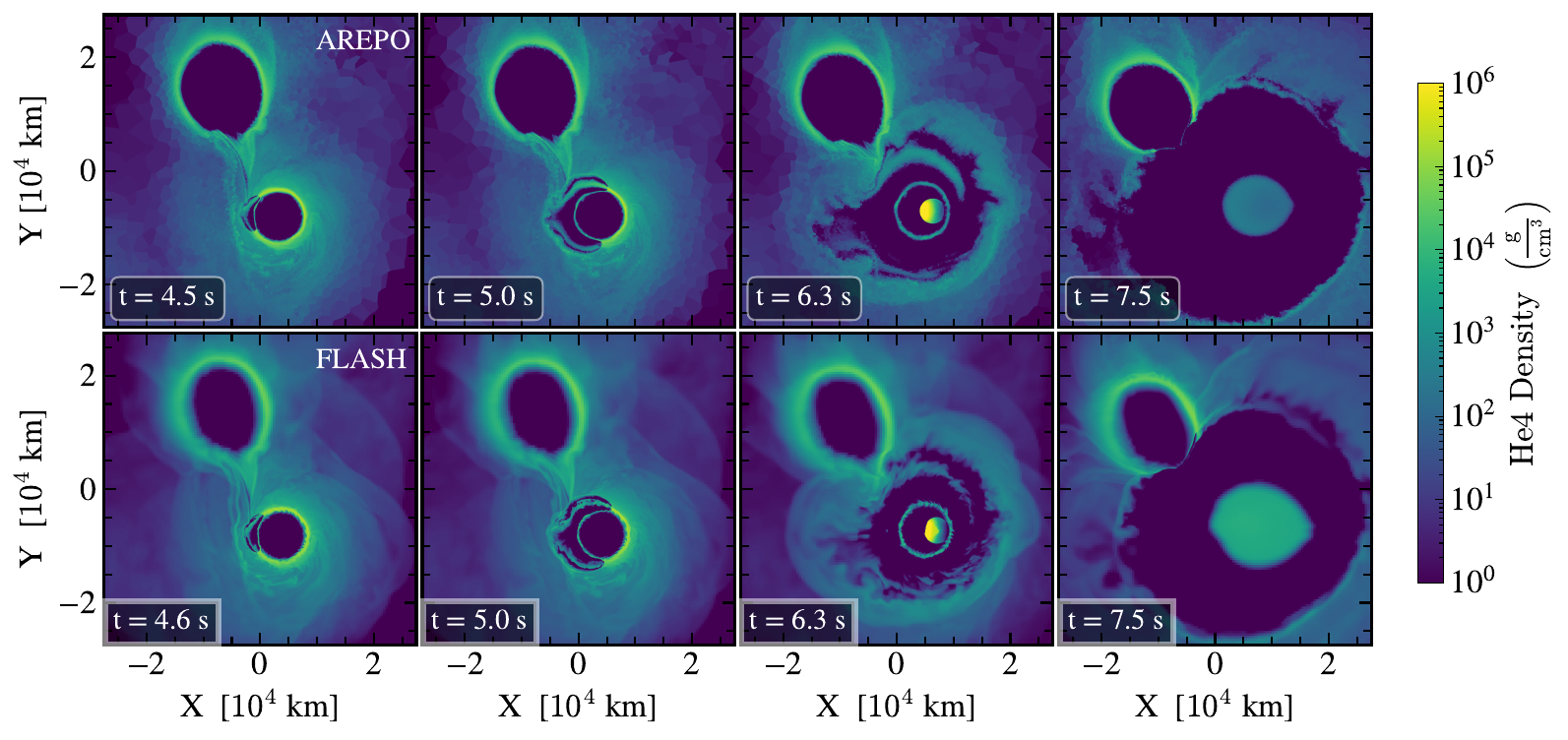}{0.95\linewidth}{(b) $^{4}\mathrm{He}$ density evolution in both AREPO (top row) and FLASH (bottom row).}}
    \caption{Time evolution of the \texttt{Double\_Det} model shown in mass density (a) and $^{4}\mathrm{He}-$density (b) fields. In each sub-figure, the top row displays the \texttt{AREPO} simulation and the bottom row displays the \texttt{FLASH} simulation at corresponding evolutionary stages. All slices are taken at the $z=0$ mid-plane of the simulation domain. The mass density field shows the density structure of the binary, and $^{4}\mathrm{He}-$density highlights helium depletion due nuclear burning. From left to right: The two columns show the early propagation and wrap-around of the helium detonation on the primary WD surface; The third column displays the core detonation initiated by shock convergence; The fourth column shows the complete disruption of the primary WD with the ejecta impacting the secondary WD.}
    \label{fig:double_det_densities_evolution}
\end{figure}

We first map the initial conditions of the \texttt{Double\_Det} model onto the computational grid of the \texttt{FLASH} code, where the subsequent evolution of the binary proceeds entirely independent of the \texttt{AREPO} simulation. The \texttt{FLASH} simulation begins from the state in which 
the Roche-lobe of the secondary WD is overfilled and matter from the secondary WD is accreted on to the primary WD. As the system evolves, the accretion stream extract helium-rich material from the secondary’s outer layers while also drawing carbon/oxygen-rich material from its inner layers. This create a stratified structure with helium encrusting the outside, while carbon/oxygen material flowing within. The accretion stream deposits this material onto the surface of the primary WD with consistent impact. This rapid impact compresses the helium layers at the impact site, generating a hot-spot while also increasing the temperature and density. In this process, the thermodynamic conditions at impact location continue to evolve toward a state in which a helium runaway can initiate \citep{guillochon2010surface, pakmor2013helium}. At approximately $t = 4.3 \ \rm{s}$ into the simulation, explosive helium burning ignites at the site of accretion stream impact located roughly $1100 \ \mathrm{km}$ in the positive $z-$axis. Figure \ref{fig:double_det_densities_evolution} displays the slice plots of the mass density (a) and the $^{4}\mathrm{He}-$density (b) from \texttt{AREPO} (top rows) and \texttt{FLASH} (bottom rows) simulations. These snapshots compare similar states from both simulation with increasing in time from left to right columns.

In order to understand the underlying nuclear reactions responsible for initiating the helium ignition, we analyze the temperature and compositional evolution of fluid elements using passive Lagrangian tracer particles. Our analysis procedure begins with identifying the tracer particle with the highest temperature on the simulation domain before and after the onset of helium ignition. During this time interval the peak temperature rises from $2 \times 10^{8}\ \rm K$ to $2 \times 10^{9}\ \rm K$. We then identify the position of this tracer on the simulation domain and use the \texttt{scipy.spatial.KDTree} algorithm \citep{friedmanetal77} to find all neighboring tracers within a $350 \rm \ km$ radius of this particle to outline a localized region where ignition has likely begun.

Next, we filter the particles to include only those where the $^{4}\rm He$ mass fraction has changed by more than $20\%$ relative to the previous snapshot of the tracer particles, taken five time-steps prior to the current snapshot. This ensures that only tracers experiencing active helium burning are considered in the analysis. For each of these selected tracers, we then track the compositional histories from the start of the simulation ($t = 0$) up to ignition, monitoring the evolution of $^{4}\rm He$, $^{12}\rm C$, $^{16}\rm O$, and $^{20}\rm Ne$ mass fractions over time. By examining the particle composition history, we determine whether the helium layer ignition was driven by the triple-$\alpha$ reaction or due to the $\alpha$-capture with heavier seed nuclei such as carbon or oxygen. If $^{4}\rm He$ decreases while $^{12}\rm C$ increases just prior to ignition, the triple-$\alpha$ reaction is responsible. Conversely, if $^{4}\rm He$ decreases along with a decrease in $^{12}\rm C$ or $^{16}\rm O$, the ignition is primarily driven by $\alpha$-capture onto heavier seed nuclei.

In our analysis, we identify two tracer particles showing the first signs of helium burning at $\rm t = 4.25 \ s$, where the mass fraction of helium ($\rm X[{^{4}\rm He}]$) has decreased from $0.2$ to below $0.1$ in the duration of temperature increase. A key observation from these particles’ compositional histories is that the abundances remain essentially unchanged until the time of ignition. During ignition, both $\rm X[{^{4}\rm He}]$ and $\rm X[{^{16}\rm O}]$ decrease simultaneously while $\rm X[{^{20} Ne]}$ has increased in both particles, suggesting that the $\alpha$-capture process onto heavier $^{16}\rm O$ nuclei has likely seeded the ignition. In the next few time steps, the temperature of surrounding fluid elements is also increases with a significant drops of $\rm X[{^{4}\rm He}]$ and $\rm X[{^{12}\rm C}]$ or $\rm X[{^{16}\rm O}]$, indicating that the burning is propagating inside the helium layer and transitions into a self-sustained helium detonation.

\subsubsection{Shock convergence and the core detonation of the primary white dwarf} 

\begin{figure}[htb!]
    
    \gridline{\fig{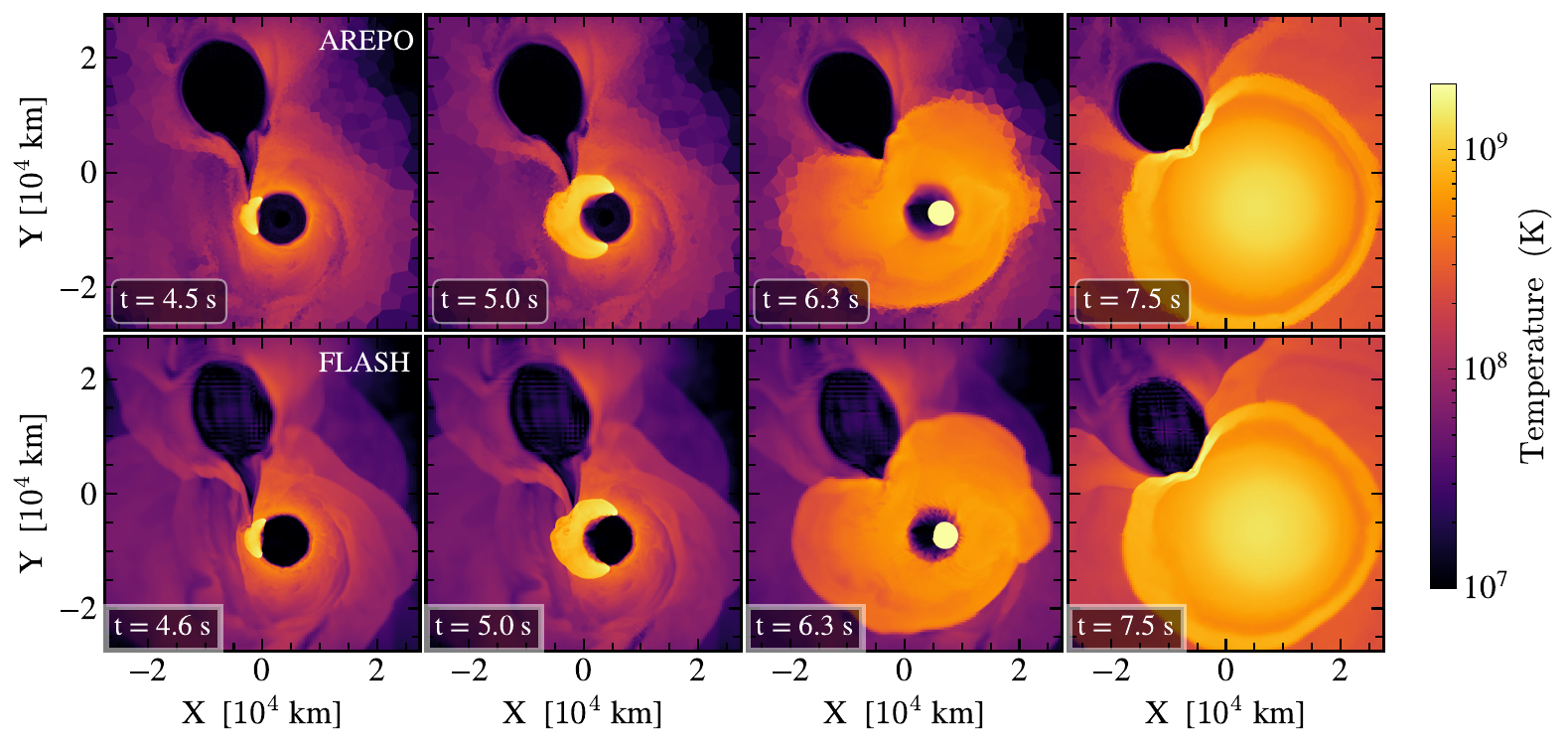}{0.95\linewidth}{(a) Temperature evolution in both AREPO (top row) and FLASH (bottom row).}}
    \gridline{\fig{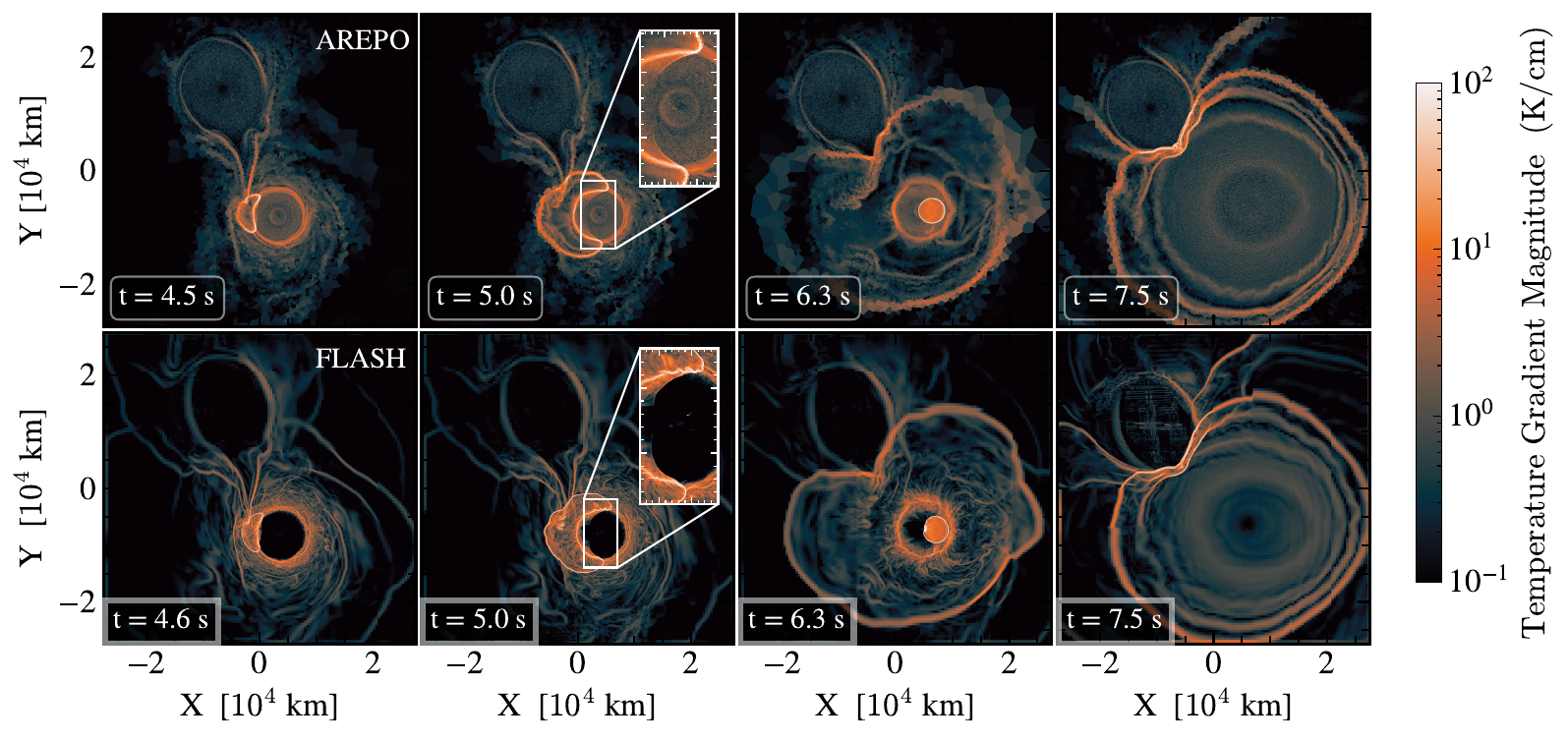}{0.95\linewidth}{(b)  Temperature gradient magnitude evolution in both AREPO (top row) and FLASH (bottom row).}}
    \caption{Time evolution of the \texttt{Double\_Det} model in the temperature (a) and the temperature gradient magnitude (b) fields. As in Figure \ref{fig:double_det_densities_evolution}, the top and bottom rows compare \texttt{AREPO} and \texttt{FLASH} simulations at corresponding evolutionary stages. All slices are taken at the $z=0$ mid-plane. The temperature field highlights regions of active nuclear burning, while the temperature gradient magnitude captures detonation fronts and the intricate internal shock structure. From left to right: the first two columns show the early surface helium detonation and its wrap-around (with zoomed-in insets revealing internal shocks); the third column displays the core detonation initiated after shock convergence; the fourth column shows the complete disruption of the primary WD and the resulting ejecta impact on the secondary WD.}
    \label{fig:double_det_temp_evolution}
\end{figure}

In the following evolution, the helium detonation fronts symmetrically begin to expand in lateral directions, as shown by the slice plots (in the first and second columns of Figures \ref{fig:double_det_densities_evolution} and \ref{fig:double_det_temp_evolution}). Meanwhile, a strong compressive shock is produced from the detonation traveling inward from the ignition location to the core of the WD. We refer to this as the primary shock. As the helium detonation rapidly slides onto the surface, it continuously transmits successive radially inward-directed compressive shocks below the outer shell of the WD. We show the internal structures of these shocks in the temperature‐gradient slices shown in the second column of Figure \ref{fig:double_det_temp_evolution}.

These shocks initially align in the direction of the primary shock but progressively reorient relative to it--first toward a perpendicular direction and eventually opposite to the primary shock--as the detonation front expands laterally. We refer to these as the secondary shocks.As the primary shock propagates towards the interior of the WD, some of these secondary shocks also get merged into the primary shock, collectively contributing to the growing compression in the central region. Approximately $1.3$ seconds after the helium layer ignition, the detonation has completely burned the surface helium layer of the primary, and the propagating detonation fronts merge at the far side of the ignition point, similar to the shown by the previous studies \citep{guillochon2010surface, moll2013multi, pakmor2013helium, roy20223d}. The secondary shocks produced by the merged detonation fronts move in the direction opposite to that of the primary shock.

At around $t = 6.08$ s, the primary and secondary shocks cross into each other near the center at roughly $550 \rm \ km$ offset in the negative direction from the mid-plane of the $z-$axis. To quantify the convergence of shocks, we track the density evolution of Lagrangian tracer particles in this region from the helium ignition to the post-shock convergence state. Our analysis shows that upon the convergence, the density of the particle sharply rises by approximately a factor of $7$ from $1.6\times 10^{7} \rm g \ cm^{-3}$  to  $1.1\times 10^{8} \rm g \ cm ^{-3}$. To interpret this density increase, we estimate the expected compression of density in the shock-normal direction of an ideal 1D strong shock and shock-shock convergence cases using the Rankine-Hugoniot relation \eqref{eq:1}. A complete step-by-step estimation for both cases is given in the appendix section \ref{appendix}.

The measured density compression factor $(\approx 7)$ in the simulation is shown to marginally exceeding the one-dimensional ideal single-shock strong limit $(\approx 5.65)$ for the adiabatic index $(\gamma = 1.43)$ of the tracer particle, but is significantly lower than the idealized 1D shock-shock convergence upper bound $(\rm factor \approx 18.8)$. These analytical estimates represent the idealized, one-dimensional limit where shocks are perfectly normal, planar, and infinitely strong. In contrast, the three-dimensional geometry in our simulation involves intricate shock structures with finite Mach numbers, which reduces the achievable compression. Consequently, the simulated density amplification of roughly $7$ is consistent with a non-ideal, shock–shock convergence in a realistic multi-dimensional environment. We note that these values correspond to our finest resolution of $13.4 \ \mathrm{km}$, and the results may vary with coarser or finer resolutions.

In the convergence region, the sudden density increase is followed by a rapid temperature rise from $\sim 10^{7}\ \mathrm{K}$ to $\sim 6\times 10^{8}\ \mathrm{K}$. When the primary shock enters the region previously compressed by the opposite-propagating secondary shock, the temperature rises again, reaching $\sim 5.5\times 10^{9}\ \rm{K}$ At this instance, the mass fraction of $^{12}\rm C$ rapidly declines, followed by a decrease in $^{16}\rm O$, indicating the onset of a thermonuclear runaway that quickly develops into a quasi-spherical core detonation. A slice plot of the simulation state $0.15\ \mathrm{s}$ after core detonation initiation is shown in the third column of Figure \ref{fig:double_det_temp_evolution}.

Within a second, the core detonation unbinds the primary WD and the resulting ejecta start to propagate radially outward. By $t = 7.5\ \mathrm{s}$, while the ejecta is still undergoing active nuclear burning, the explosion reaches and impacts the nearby secondary WD (see the fourth column of Figures \ref{fig:double_det_densities_evolution} and \ref{fig:double_det_temp_evolution}). Part of the ejecta approaching the secondary WD erode its outer layers while also depositing burning products onto the hemisphere facing the explosion. Due to this ablating and mixing of helium with the burning isotopes, the $\alpha-$capture processes ignite the helium layer on the secondary WD. While this burning produces the helium detonation, the less dense helium on the surface makes it asymmetric. The shocks generated from this detonation are weaker and converge in the outer layers of star, not causing a core detonation. We evolve the \texttt{Double\_Det} model in \texttt{FLASH} up to a maximum simulation time of $t_{\rm max} = 14 $ s. By then the nuclear energy generation rate falls below $ 10^{40} \ \rm erg \ s^{-1}$ while the changes in overall abundances on the domain are negligible.

In the \texttt{FLASH} simulation, the computational domain spans a total of $55,000 \ \mathrm{km}$ in each spatial direction which is sufficiently large to capture the evolution of binary over a few dynamical times. However, after the detonation of primary WD, the unbound material is expelled at nearly $10,000 \ \mathrm{km \ s^{-1} }$. As a result, a little over a second after the core detonation of the primary WD, the ejecta begins to leave the computational domain while it is still undergoing active nuclear burning.

\begin{table}[htbp]
\centering
\small 
\setlength{\tabcolsep}{6pt} 
\renewcommand{\arraystretch}{1.1} 
\begin{tabular}{lccc}
\toprule
\textbf{Species} & \textbf{Units} &\textbf{FLASH} &  \textbf{AREPO} \\
\midrule
${}^{4}\mathrm{He}$  & $\mathrm{M}_{\odot}$ & $1.124 \times 10^{-2}$  & $9.459 \times 10^{-3}$ \\
${}^{12}\mathrm{C}$  & $\mathrm{M}_{\odot}$ & $3.055 \times 10^{-1}$  & $3.073 \times 10^{-1}$ \\
${}^{16}\mathrm{O}$  & $\mathrm{M}_{\odot}$ & $3.645 \times 10^{-1}$  & $3.840 \times 10^{-1}$ \\
${}^{20}\mathrm{Ne}$ & $\mathrm{M}_{\odot}$ & $6.090 \times 10^{-3}$  & $5.592 \times 10^{-3}$ \\
${}^{24}\mathrm{Mg}$ & $\mathrm{M}_{\odot}$ & $1.283 \times 10^{-2}$  & $2.042 \times 10^{-2}$ \\
${}^{28}\mathrm{Si}$ & $\mathrm{M}_{\odot}$ & $2.657 \times 10^{-1}$  & $2.668 \times 10^{-1}$ \\
${}^{32}\mathrm{S}$  & $\mathrm{M}_{\odot}$ & $1.392 \times 10^{-1}$  & $1.967 \times 10^{-1}$ \\
${}^{36}\mathrm{Ar}$ & $\mathrm{M}_{\odot}$ & $2.937 \times 10^{-2}$  & $6.250 \times 10^{-2}$ \\
${}^{40}\mathrm{Ca}$ & $\mathrm{M}_{\odot}$ & $2.948 \times 10^{-2}$  & $7.632 \times 10^{-2}$ \\
${}^{44}\mathrm{Ti}$ & $\mathrm{M}_{\odot}$ & $8.483 \times 10^{-4}$  & $2.915 \times 10^{-3}$ \\
${}^{48}\mathrm{Cr}$ & $\mathrm{M}_{\odot}$ & $3.187 \times 10^{-4}$  & $4.259 \times 10^{-3}$ \\
${}^{52}\mathrm{Fe}$ & $\mathrm{M}_{\odot}$ & $6.388 \times 10^{-3}$  & $2.091 \times 10^{-2}$ \\
${}^{56}\mathrm{Ni}$ & $\mathrm{M}_{\odot}$ & $4.305 \times 10^{-1}$  & $2.504 \times 10^{-1}$ \\
\hline
$\Delta \ \mathrm{E_{nuc}}$   & $\mathrm{erg}$       & $1.315 \times 10^{51}$ & $1.258 \times 10^{51}$ \\
$\rm M_{ej}$                  & $\mathrm{M}_{\odot}$   & $1.001$                & $1.003$                 \\
\hline
\end{tabular}
\caption{Comparison of isotopic mass abundances, total nuclear energy release, and total ejecta mass between the \texttt{FLASH} and \texttt{AREPO} simulations for the \texttt{Double\_Det} model. Both simulations are matched at the similar evolutionary phase (see fourth column in Figures \ref{fig:double_det_densities_evolution} and \ref{fig:double_det_temp_evolution}) shortly after the core detonation of the primary WD. Values represent phase-matched, not final, nucleosynthetic abundances..}
\label{tab:double_det_ejecta_comparison}
\end{table}

Because of this limitation, we restrict our analysis to an evolutionary phase in which \texttt{FLASH} simulation still retains the more than $99 \%$ of the initial mass on the computational domain. We then compare this stage to the corresponding \texttt{AREPO} snapshot, evaluated over the identical computational volume. Table \ref{tab:double_det_ejecta_comparison} shows comparison of the mass of each species in \texttt{FLASH} and \texttt{AREPO} simulations at $\mathrm{t=7.5 \ s}$. 

While \texttt{AREPO} and \texttt{FLASH} simulations show agreement in the abundances of $^{4}\mathrm{He}, ^{12}\mathrm{C}, ^{16}\mathrm{O}, $ and $ ^{20}\mathrm{Ne}$, the differences are significant ($ > 20 \%$) for several heavier isotopes such as $^{24}\mathrm{Mg}$ and from  $^{32}\mathrm{S}$ to $^{56}\mathrm{Ni}$. In particular, the \texttt{FLASH} simulation produces roughly twice as much $^{56}\mathrm{Ni}$ as \texttt{AREPO} at the same evolutionary phase. This discrepancy can be explained by the differences in the nuclear reaction networks used by each codes in these simulations. The $13-$isotope network used by \texttt{AREPO} and the \texttt{approx13} network used by \texttt{FLASH} both employ similar $\alpha$-chain reaction rates to model nuclear burning. However, the \texttt{approx13} network additionally includes side-chain reaction rates for intermediate isotopes (e.g., $^{27}\mathrm{Al}$, $^{31}\mathrm{P}$, $^{35}\mathrm{Cl}$, up to $^{55}\mathrm{Co}$), which establish equilibrium more rapidly at temperatures above $3\times10^{9}\ \mathrm{K}$. This provides additional and faster burning pathways that lead to increased production of $^{56}\mathrm{Ni}$ compared to a network with only the standard $\alpha$-chain. 

\subsection{{\rm \texttt{Quad\_Det}} model} \label{subsec:quad_det}

\subsubsection{Helium layer ignition of the primary white dwarf}

\begin{figure}[htb!]
    
    \gridline{\fig{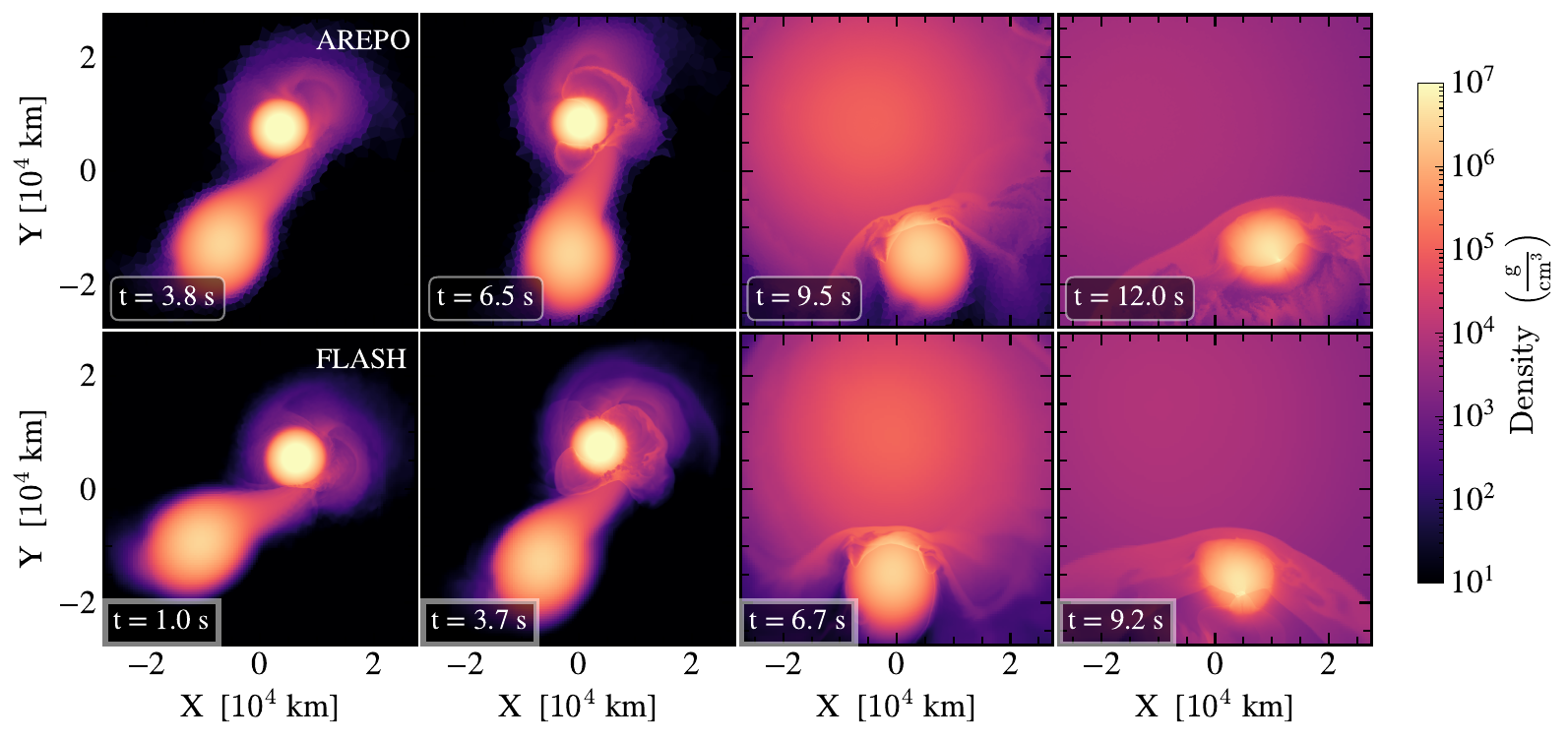}{0.95\linewidth}{(a) Mass density evolution in both AREPO (top row) and FLASH (bottom row).}}
    \gridline{\fig{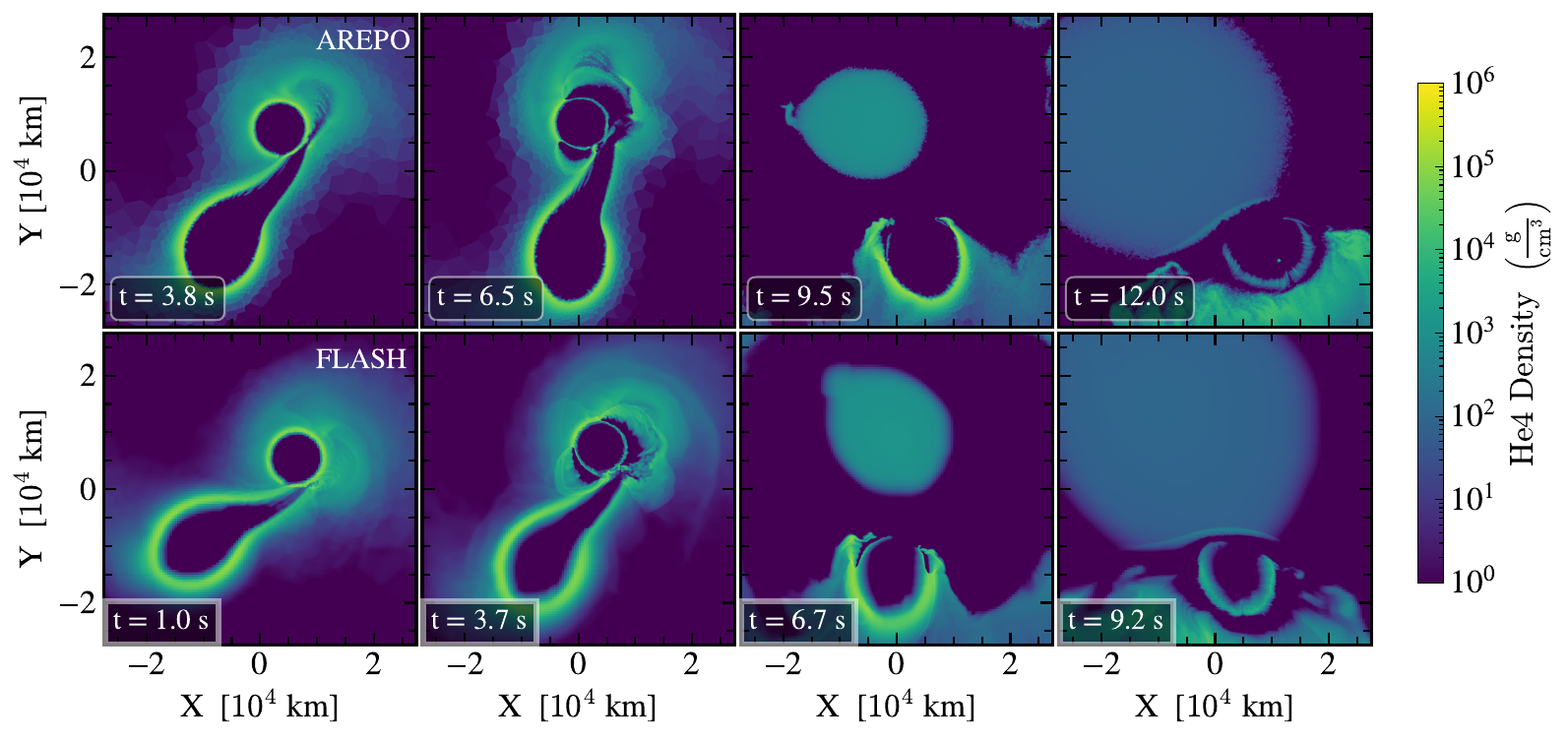}{0.95\linewidth}{(b) $^{4}\mathrm{He}$ density evolution in both AREPO (top row) and FLASH (bottom row).}}
    \caption{Time evolution of the \texttt{Quad\_Det} model shown in mass density (a) and $^{4}\mathrm{He}-$density (b) fields. In each sub-figure, the top row displays the \texttt{AREPO} simulation and the bottom row displays the \texttt{FLASH} simulation at corresponding evolutionary stages. Because helium ignition in \texttt{FLASH} occurs $\sim 2.8\text{ s}$ earlier than in \texttt{AREPO}, the rows are phase-matched to the onset of surface ignition rather than absolute simulation time. All slices are taken at the $z=0$ mid-plane of the simulation domain. The mass density field shows the density structure of the binary, and $^{4}\mathrm{He}-$density highlights helium depletion due to nuclear burning. From left to right: the first column shows the pre-detonation state approximately $1.4\text{ s}$ prior to surface helium ignition of the primary WD; the second column illustrates the lateral propagation of the helium detonation across the primary WD surface ($\sim 1.3\text{ s}$ post-ignition); the third column shows the ejecta from the exploded primary impacting the secondary WD ($\sim 2\text{ s}$ after primary core detonation), and igniting its surface helium detonation; the fourth column shows the initiation of the core detonation in the secondary WD, by the convergence of shocks. }
    \label{fig:quad_det_densities_evolution}
\end{figure}

Similar to the initial state of the \texttt{Double\_Det} model, the \texttt{FLASH} simulation of the \texttt{Quad\_Det} model also starts from a state in which the Roche lobe filling secondary WD is channeling helium-rich matter onto the surface of the primary WD through a dense accretion stream. The continuous impact from the accretion stream rapidly compresses the outer shells at the impact location and creates a hot-spot. At approximately $\mathrm{t = 2.4 \ s}$ in the simulation time, helium is ignited at the impact site located roughly $300 \ \mathrm{km}$ from the mid-plane. To identify the nuclear reactions responsible for this helium ignition, we apply the same Lagrangian tracer-based analysis methodology used in the \texttt{Double\_Det} model. We identify the tracer where helium ignition was began by inspecting the evolution of the maximum temperature of the tracers prior to the onset of helium burning. After identifying this tracer, we examine its composition history to determine whether the initial mass fractions of this tracer were significantly changed prior to the onset of helium ignition.

The compositional history of this particle prior to the rapid temperature increase shows the initial mass fraction of the $\mathrm{^{4}He}$ was substantially decreased from $0.9$ to $0.4$ while the $\mathrm{^{12}C}$ mass fraction was increased from $0.09$ to $0.3$. During this time interval, the temperature remained roughly $\mathrm{6\times10^{8} \ K}$. The changes in the composition of the particle clearly imply the occurrence of an active triple-$\alpha$ reaction. In the same duration, a simultaneous increase of $\mathrm{^{16}O}$ to $\sim0.25$ indicates that the some carbon is also consumed by the subsequent $\alpha$-capture.  At $\mathrm{t\approx2.4 \ s}$, the temperature increases to $\mathrm{1.3 \times 10^{9} \ K}$, which accelerated the reaction rate for $\alpha$-capture on $\mathrm{^{16}O}$ forming $\mathrm{^{20}Ne}$ while consuming a large fraction of remaining $\mathrm{^{4}He}$. This marks the runaway burning of helium, which begins to spread away from the ignition site to adjacent fluid elements while also increasing the temperature to $\mathrm{1.6 \times 10^{9} \ K}$. In the subsequent time steps, this ignition transitions to a helium detonation which propagates laterally away from the ignition point, similar to the \texttt{Double\_Det} model.

In comparison to \texttt{Double\_Det} models evolved in \texttt{FLASH} and \texttt{AREPO} in which the helium ignition occur approximately at the same time, the ignition times for \texttt{Quad\_Det} models in both codes vary significantly. In \texttt{FLASH} simulation of the \texttt{Quad\_Det} model, the helium layer was ignited at approximately $2.4\ \mathrm{s}$, which was about $2.8\ \mathrm{s}$ earlier than the ignition time in the \texttt{AREPO} simulation. Figure \ref{fig:quad_det_densities_evolution} shows the mid-plane mass density and density of $\mathrm{^{4}He}$ in the \texttt{Quad\_Det} model at different evolutionary phases (increasing in time from left to right), while also comparing identical phases from \texttt{AREPO} (top rows) and \texttt{FLASH} (bottom rows). These phases are synchronized by the approximate time of onset of helium layer ignition on the surface of the primary WD.

\subsubsection{Shock convergence and the core detonation of the primary white dwarf} 

\begin{figure}[htb!]
    
    \gridline{\fig{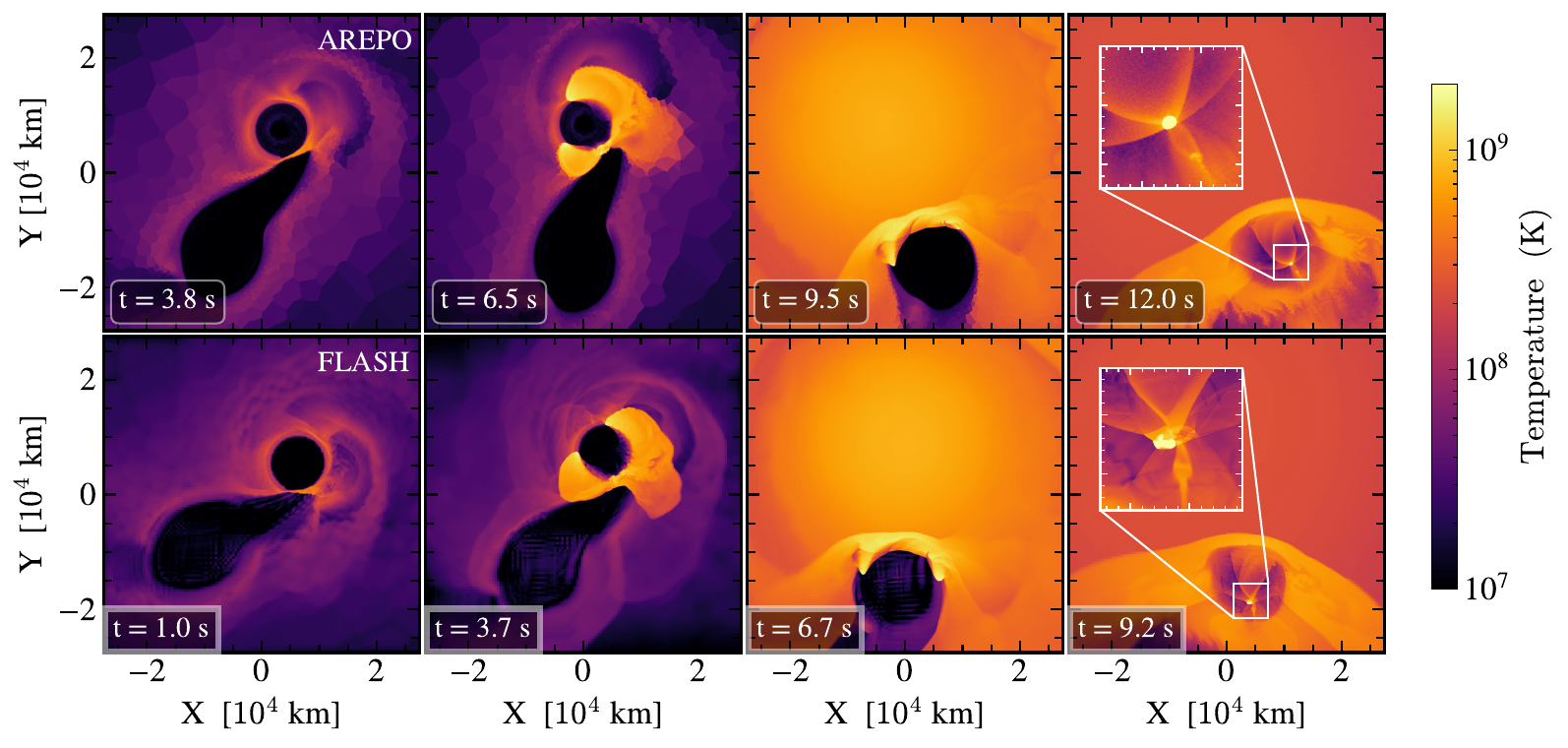}{0.95\linewidth}{(a) Temperature evolution in both AREPO (top row) and FLASH (bottom row).}}
    \gridline{\fig{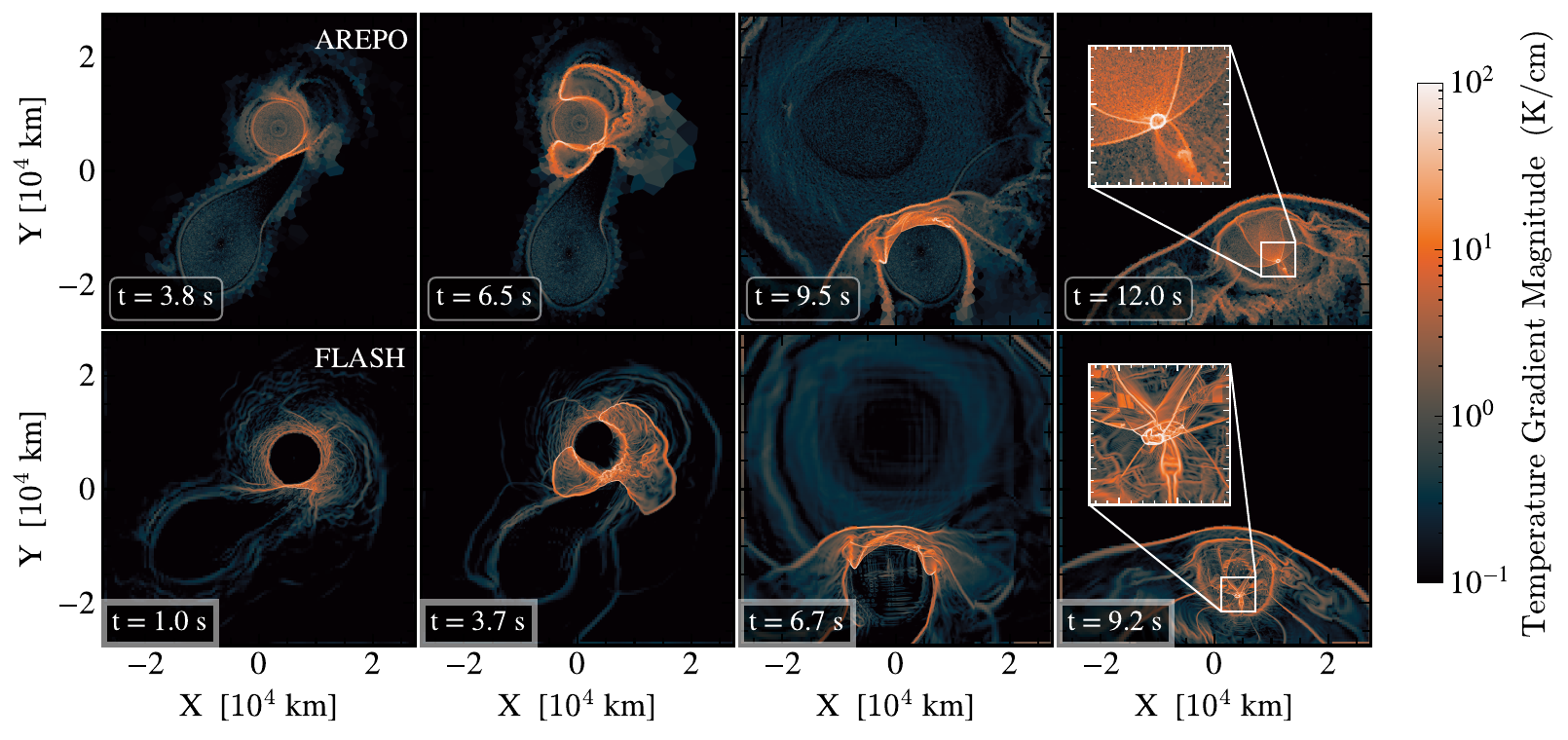}{0.95\linewidth}{(b)  Temperature gradient magnitude evolution in both AREPO (top row) and FLASH (bottom row).}}
    \caption{Time evolution of the \texttt{Quad\_Det} model shown in temperature (a) and temperature gradient magnitude (b) fields. As in Figure 3, the \texttt{AREPO} (top row) and \texttt{FLASH} (bottom row) simulations are phase-matched to account for the $\sim 2.8\text{ s}$ ignition time offset between the simulations. The temperature field highlights regions with higher temperatures and nuclear burning, while the temperature gradient magnitude captures the structure of the detonation fronts and shock interactions. From left to right: the first column shows the pre-detonation state approximately $1.4\text{ s}$ prior to surface helium ignition of the primary WD; the second column highlights the lateral propagation of the helium detonation and the resulting internal shocks ($\sim 1.3\text{ s}$ post-ignition); the third column captures the ejecta from the exploded primary impacting the secondary WD ($\sim 2\text{ s}$ after primary core detonation), igniting helium on the surface of the secondary WD; the fourth column shows the shocks produced by helium detonation, converge near the core and the igniting the core detonation in the secondary WD (zoomed-in inset plots show the core detonation initiated by the shock convergence).}
    \label{fig:quad_det_temp_evolution}
\end{figure}

As the helium detonation propagates on the surface of the primary WD, it constantly produces compressive shocks below the detonation fronts that travel from the outer layers to the core of the primary WD. During the shock propagation, they produce a similar spatial structure as we have previously described in the \texttt{Double\_Det} model. The primary shock starts from the ignition location, meanwhile the secondary shocks are also produced from the laterally propagating detonation fronts. These secondary shocks travel in radially inward directions, while also re-orienting relative to the primary shock. At approximately $\mathrm{t = 4.2 \ s}$, the helium detonation completely wraps around the WD while the detonation fronts merge at the far side of the helium layer ignition site. The final shock produced by this merge, travels towards the core from the opposite direction of the primary shock. 

Around $t \approx 4.6$ s, the primary and secondary shocks converge near the center of the WD at roughly, producing a sharp increase in density from $\mathrm{1.15\times10^{7}\ g\ cm^{-3}}$ to $\mathrm{5.73\times10^{7}\ g\ cm^{-3}}$ by a factor of $\approx 5$. For comparison, the maximum compression factor for an ideal one-dimensional single strong shock predicted by the Rankine–Hugoniot relation (Eq.\ref{eq:1}) of $\approx 5.5$ for an adiabatic index of $1.42$ for this tracer. Unlike the \texttt{Double\_Det} model, where we found a higher compression factor of $\approx 7$ (slightly above the single strong shock limit), in \texttt{Quad\_Det} model, this compression factor is slightly lower than the ideal one-dimensional strong shock limit. 

The primary reason for this discrepancy is closely related to the kinetic energy released by the helium detonation on the primary WD. In the \texttt{Double\_Det} model, the primary WD has a helium layer of $\approx 0.01 \ \mathrm{M_\odot}$, making it denser and more compact, whereas in the \texttt{Quad\_Det} model, the primary WD carries much lower $\approx 0.005 \ \mathrm{M_\odot}$ of helium on its surface. The lower-mass helium shell on the primary in the \texttt{Quad\_Det} model burns at lower densities, producing slower detonation speeds and less intense shocks. This also lengthens the burning duration for helium detonation, taking $\sim1.8$ s compared to $\sim1.3$ s in the \texttt{Double\_Det} model. As a result, the converging shocks near the WD center are slightly weaker, leading to a lower density amplification. However, the measured compression factor in our simulation is consistent with the compression expected from shocks with finite Mach numbers undergoing non-ideal, three-dimensional shock–shock convergence. The values we reported in this analysis correspond to our finest resolution $13.4 \mathrm{\ km}$, and can vary with the change in resolution.

Despite the lower compression, this density rise is immediately followed by a rapid temperature increase from $\sim 10^{7}\ \mathrm{K}$ to $\sim 6\times10^{9}\ \mathrm{K}$, rapidly heating the gas which triggers the core detonation of the primary WD. From this point onward, the core detonation propagates from the center outward, completely disrupting the primary WD within the next $\sim 0.6$ seconds. The unbound mass from the explosion is expelled as nearly spherical ejecta, while the expanding ejecta simultaneously burns through the incoming accretion stream.

\subsubsection{Helium layer detonation and core detonation of the secondary white dwarf}

By approximately $6.0\ \mathrm{s}$, the expanding ejecta consume the accretion stream and reach the secondary WD, impacting its helium-rich outer shell. Because the ejecta are still undergoing active nuclear burning, upon impact they quickly trigger helium ignition on the surface of the secondary WD. Similar to the ignition of helium layer on the secondary WD previously described for the \texttt{Double\_Det} model, the high temperatures of the impacting ash $(\gtrsim 10^{9}\ \mathrm{K})$ facilitate efficient $\alpha$-capture processes onto partially burned IMEs. This helium ignition quickly transitions into a detonation that begins to propagate from the impact site toward the antipode. We show the propagation of the helium detonation on the surface of the secondary WD at $\mathrm{t=6.7\ s}$ in the \texttt{FLASH} simulation and $\mathrm{t=9.5\ s}$ in the \texttt{AREPO} simulation in the third column of Figures \ref{fig:quad_det_densities_evolution} and \ref{fig:quad_det_temp_evolution}. These snapshots correspond to the same evolutionary state between the two simulations, accounting for the $2.8\ \mathrm{s}$ temporal offset. 
\begin{table}[ht]
\centering
\begin{minipage}{0.48\textwidth}
\centering
\small 
\setlength{\tabcolsep}{6pt} 
\renewcommand{\arraystretch}{1.1} 
\begin{tabular}{lccc}
\toprule
\textbf{Species} & \textbf{Units} &\textbf{FLASH} &  \textbf{AREPO} \\
\midrule
${}^{1}\mathrm{H}$   & $\mathrm{M}_{\odot}$ &         --              &         --           \\
${}^{3}\mathrm{He}$  & $\mathrm{M}_{\odot}$ &         --              &         --           \\
${}^{4}\mathrm{He} $ & $\mathrm{M}_{\odot}$ & $2.475 \times 10^{-2}$  & $2.406 \times 10^{-2}$ \\
${}^{11}\mathrm{B}$  & $\mathrm{M}_{\odot}$ &         --              &         --             \\
${}^{12}\mathrm{C}$  & $\mathrm{M}_{\odot}$ & $2.929 \times 10^{-1}$  & $2.892 \times 10^{-1}$ \\
${}^{13}\mathrm{C} $ & $\mathrm{M}_{\odot}$ &         --            &           --           \\
${}^{13}\mathrm{N}$  & $\mathrm{M}_{\odot}$ &         --            &           --           \\
${}^{14}\mathrm{N} $ & $\mathrm{M}_{\odot}$ &         --            &           --           \\
${}^{15}\mathrm{N}$  & $\mathrm{M}_{\odot}$ &         --            &           --           \\
${}^{15}\mathrm{O}$  & $\mathrm{M}_{\odot}$ &         --            & $1.397 \times 10^{-6}$ \\
${}^{16}\mathrm{O} $ & $\mathrm{M}_{\odot}$ & $3.411 \times 10^{-1}$ & $3.514 \times 10^{-1}$ \\
${}^{17}\mathrm{O} $ & $\mathrm{M}_{\odot}$ &         --            &           --           \\
${}^{18}\mathrm{F}$  & $\mathrm{M}_{\odot}$ &         --            &           --           \\
${}^{19}\mathrm{Ne}$ & $\mathrm{M}_{\odot}$ &         --            &           --           \\
${}^{20}\mathrm{Ne}$ & $\mathrm{M}_{\odot}$ & $6.680 \times 10^{-3}$  & $6.557 \times 10^{-3}$ \\
${}^{21}\mathrm{Ne}$ & $\mathrm{M}_{\odot}$ &         --            &           --           \\
${}^{22}\mathrm{Ne} $& $\mathrm{M}_{\odot}$ &         --            &           --           \\
${}^{22}\mathrm{Na}$ & $\mathrm{M}_{\odot}$ &         --            &           --           \\
${}^{23}\mathrm{Na}$ & $\mathrm{M}_{\odot}$ &         --            & $2.112 \times 10^{-6}$ \\
${}^{23}\mathrm{Mg} $& $\mathrm{M}_{\odot}$ &         --            & $9.480 \times 10^{-6}$ \\
${}^{24}\mathrm{Mg}$ & $\mathrm{M}_{\odot}$ & $1.211 \times 10^{-2}$  & $1.006 \times 10^{-2}$ \\
${}^{25}\mathrm{Mg}$ & $\mathrm{M}_{\odot}$ &         --            & $7.185 \times 10^{-7}$ \\
${}^{26}\mathrm{Mg}$ & $\mathrm{M}_{\odot}$ &         --            & $4.054 \times 10^{-7}$ \\
${}^{25}\mathrm{Al}$ & $\mathrm{M}_{\odot}$ &         --            & $1.866 \times 10^{-6}$ \\
${}^{26}\mathrm{Al}$ & $\mathrm{M}_{\odot}$ &         --            & $2.234 \times 10^{-5}$ \\
${}^{27}\mathrm{Al}$ & $\mathrm{M}_{\odot}$ &         --            & $1.788 \times 10^{-5}$ \\
${}^{28}\mathrm{Si}$ & $\mathrm{M}_{\odot}$ & $2.636 \times 10^{-1}$ & $1.647 \times 10^{-1}$ \\
${}^{29}\mathrm{Si}$ & $\mathrm{M}_{\odot}$ &         --            & $7.173 \times 10^{-5}$ \\
\hline
$\mathrm{M_{ej}}$   & $\mathrm{M}_{\odot}$  &  $0.934$  &  $0.912$ \\
\hline
\end{tabular}
\end{minipage}%
\hfill
\begin{minipage}{0.48\textwidth}
\centering
\small 
\setlength{\tabcolsep}{6pt} 
\renewcommand{\arraystretch}{1.1} 
\begin{tabular}{lccc}
\toprule
\textbf{Species} & \textbf{Units} &\textbf{FLASH} &  \textbf{AREPO} \\
\midrule
${}^{30}\mathrm{Si}$ & $\mathrm{M}_{\odot}$ &         --            & $5.023 \times 10^{-6}$ \\
${}^{29}\mathrm{P}$  & $\mathrm{M}_{\odot} $&         --            & $2.044 \times 10^{-4}$ \\
${}^{30}\mathrm{P}$ & $\mathrm{M}_{\odot} $&         --            & $1.405 \times 10^{-5}$ \\
${}^{31}\mathrm{P}$  & $\mathrm{M}_{\odot}$ &         --            & $3.942 \times 10^{-5}$ \\
${}^{31}\mathrm{S} $ & $\mathrm{M}_{\odot}$ &         --            & $7.852 \times 10^{-5}$ \\
${}^{32}\mathrm{S}$  & $\mathrm{M}_{\odot}$ & $1.418 \times 10^{-1}$  & $1.430 \times 10^{-1}$ \\
${}^{33}\mathrm{S}$  & $\mathrm{M}_{\odot} $&         --            & $5.009 \times 10^{-5}$ \\
${}^{33}\mathrm{Cl}$ & $\mathrm{M}_{\odot}$ &         --            & $1.481 \times 10^{-5}$ \\
${}^{34}\mathrm{Cl}$ & $\mathrm{M}_{\odot}$ &         --            & $2.665 \times 10^{-5}$ \\
${}^{35}\mathrm{Cl} $& $\mathrm{M}_{\odot}$ &         --            & $1.233 \times 10^{-5}$ \\
${}^{36}\mathrm{Ar}$ & $\mathrm{M}_{\odot}$ & $3.093 \times 10^{-2}$  & $3.797 \times 10^{-2}$ \\
${}^{37}\mathrm{Ar}$ & $\mathrm{M}_{\odot}$ &         --            & $4.247 \times 10^{-5}$ \\
${}^{38}\mathrm{Ar} $& $\mathrm{M}_{\odot}$ &         --            & $1.308 \times 10^{-7}$ \\
${}^{39}\mathrm{Ar}$ & $\mathrm{M}_{\odot}$ &         --            &          --            \\
${}^{39}\mathrm{K} $ & $\mathrm{M}_{\odot}$ &         --            & $1.696 \times 10^{-5}$ \\
${}^{40}\mathrm{Ca}$ & $\mathrm{M}_{\odot}$ & $2.813 \times 10^{-2}$  & $4.300 \times 10^{-2}$ \\
${}^{43}\mathrm{Sc}$ & $\mathrm{M}_{\odot}$ &         --            &          --              \\
${}^{44}\mathrm{Ti}$ & $\mathrm{M}_{\odot}$ & $2.475 \times 10^{-4}$  & $1.607 \times 10^{-4}$ \\
${}^{47}\mathrm{V} $ & $\mathrm{M}_{\odot}$ &         --            &          --             \\
${}^{48}\mathrm{Cr} $& $\mathrm{M}_{\odot}$ & $3.578 \times 10^{-4}$  & $9.216 \times 10^{-4}$ \\
${}^{51}\mathrm{Mn}$ & $\mathrm{M}_{\odot}$ &         --            & $4.204 \times 10^{-7}$ \\
${}^{52}\mathrm{Fe}$ & $ \mathrm{M}_{\odot}$ & $7.840 \times 10^{-3}$  & $1.700 \times 10^{-2}$ \\
${}^{54}\mathrm{Fe}$ & $\mathrm{M}_{\odot}$ & $2.179 \times 10^{-4}$  &         --            \\
${}^{56}\mathrm{Fe}$ & $\mathrm{M}_{\odot}$ &         --            & $6.762 \times 10^{-6}$ \\
${}^{55}\mathrm{Co}$ & $\mathrm{M}_{\odot}$ &         --            & $1.257 \times 10^{-5}$ \\
${}^{56}\mathrm{Ni}$ & $\mathrm{M}_{\odot}$ & $4.517 \times 10^{-1}$  & $5.155 \times 10^{-1}$ \\
${}^{58}\mathrm{Ni}$ & $\mathrm{M}_{\odot}$ &         --            & $3.267 \times 10^{-3}$ \\
${}^{59}\mathrm{Ni}$ & $\mathrm{M}_{\odot}$ &         --            & $5.579 \times 10^{-5}$ \\
\hline
$\Delta \ \mathrm{E_{nuc}}$   & $\mathrm{erg} $      & $1.353 \times 10^{51}$ & $1.386 \times 10^{51}$ \\
\hline
\end{tabular}
\end{minipage}
\caption{Comparison of isotopic mass abundances, total nuclear energy release, and total ejecta mass between the \texttt{FLASH} and \texttt{AREPO} simulations for the \texttt{Quad\_Det} model. Both simulations are matched at the similar evolutionary phase (see third column in Figures \ref{fig:quad_det_densities_evolution} and \ref{fig:quad_det_temp_evolution}) shortly after the core detonation of the primary WD. Values represent phase-matched, not final, nucleosynthetic abundances.}
\label{tab:quad_det_ejecta_comparison}
\end{table}

At this stage, we calculate the total nucleosynthetic abundances produced by the core detonation of the primary WD. To ensure a consistent comparison, both \texttt{FLASH} and \texttt{AREPO} abundances are evaluated within an identical computational domain of $55,000\ \mathrm{km}$ in each direction. We note that this domain size considered for the comparison between both simulations retains more than $90\%$ of the total initial mass to make the comparison consistent. The resulting abundances, total ejecta masses, and released nuclear energies are reported in Table \ref{tab:quad_det_ejecta_comparison}.

In comparison to the \texttt{FLASH} simulation, which uses the \texttt{approx19} nuclear reaction network, the \texttt{AREPO} simulation employs a more extensive $55-$isotope network. Consequently, a direct species-to-species comparison between the two simulations is restricted to the mutual isotopes present in both networks. Despite the differing network sizes, the production of these common IMEs, with an exception of $^{28}\mathrm{Si}$, shows excellent agreement between the two simulations. Furthermore, the abundances of mutual IGEs, such as $^{56}\mathrm{Ni}$, $^{52}\mathrm{Fe}$, and $^{48}\mathrm{Cr}$, are also in close agreement across both codes. Apart from the isotopic yields, the total ejecta masses and the released nuclear energy in both simulations are also shows an excellent agreement. 

In the subsequent evolution of the system, as the helium detonation races across the surface, it produces compressive shocks that travel inward toward the center of the secondary WD, similar the helium detonation of the primary WD. By $\mathrm{8.4}$ s, the detonation as swept across the far side, and the helium layer is completely consumed. Meanwhile, the shock produced from this merging detonation fronts moves in opposite directions of the primary shock, and converge into primary shock at $\mathrm{t = 9.22}$ s. At the location of shock convergence, the density of fluid undergo sharp increase from $10^{6} \ \mathrm{g \ cm^{-3}}$ to $8 \times 10^{6} \ \mathrm{g \ cm^{-3}}$, with a compression factor of $\approx 8$. For an ideal strong 1D shock, the predicted compression ratio is $\approx 4.5$, while the upper limit for shock-shock convergence in 1D, is predicted to be $\approx 12.1 $ for $\gamma \sim 1.58$ of the tracer. However, the shocks generated in our hydrodynamical simulations have 3D planar geometry with a bounded Mach number and finite resolution per cell, which impacts the overall compression observed. Therefore, the density compression factor of $\approx 8$ observed in our simulation is consistent with non-ideal compression from the shock convergence with finite Mach number in 3D geometry which is above the single 1D strong shock limit but below the predicted limit from the convergence of two ideal strong shocks in 1D. At the same instance, the temperature also increases from $10^{7} \ \mathrm{K}$ to $3 \times 10^{9} \ \mathrm{K}$, igniting the core detonation of the secondary WD. Subsequently, the core detonation propagate outwards exploding the secondary WD completely. We follow the evolution of the core detonation of the secondary up to $\mathrm{t = 11 \ s}$ in the simulation time, by which the nuclear energy generation rate falls below $\mathrm{10^{46} \ erg \ s^{-1}}$.

Because the secondary WD is less massive than the primary, its core detonation occur at lower density $\approx 6 \times 10^{6} \ \mathrm{g \ cm^{-3}}$. Consequently the explosion of secondary WD is less energetic releasing additional $ \approx 3.1 \times 10^{50} \ \mathrm{erg}$ of nuclear energy. By this stage, we lose significant amount of the ejecta masses from the \texttt{FLASH} simulation domain, thus we do not quantify any nucleosynthetic abundances from the core detonation of the secondary WD. 

\section{Discussion} \label{sec:discussion}

Previous studies have demonstrated that converging shocks from the detonation of the surface helium layer creates favorable conditions for carbon ignition near the locus of convergence. \citep{fink2007double, fink2010double, moll2013multi, shen2014ignition}. However, several complete three-dimensional hydrodynamical simulations of the helium-ignited binary WD merger channel have not reached a consistent conclusion regarding the detonation outcome of this channel \citep{guillochon2010surface, pakmor2013helium, roy20223d, pakmor2022fate}. While they obtained the detonation of the surface helium layers, the majority of them did not capture the predicted core detonation of the WD. This is not entirely unexpected, as the carbon detonation forms on centimeter scales that are not resolved in either the previous studies or our own simulations. Consequently, the ignition outcome can be sensitive to the numerical scheme and resolution employed.

For example, \cite{guillochon2010surface} simulated helium accreting binary WD models using a hybrid approach by combining the SPH method \citep{dan_etal_2011} to create initial conditions with the Eulerian mesh code \texttt{FLASH} \citep{fryxell2000flash} to evolve the detonation phase and later evolution. Their simulations with spatial the resolution $\gtrsim 34 \ \mathrm{km}$ successfully demonstrated the helium layer detonation, but insufficient to produce the following core detonation. In following studies, \cite{pakmor2013helium} and \cite{roy20223d}  also used a similar hybrid approach using different codes, \texttt{AREPO} and \texttt{FLASH} respectively, with a higher spatial resolution $(\sim 13-17 \ \mathrm{km})$. Both studies obtain the surface helium detonation but fail to ignite the core detonation of the primary WD. In independent, unpublished tests, the initial conditions of \cite{roy20223d} were evolved using the \texttt{AREPO} code with a similar configurations and likewise did not produce a core detonation. 

In contrast, \cite{pakmor2022fate} undertook the 3D evolution of a helium accreting binary WD model from end-to-end in the \texttt{AREPO} code with a spatial resolution $(10^{-7} \ \mathrm{M_{\odot}} \approx 10 \ \mathrm{km})$, and successfully captured both surface helium detonation and primary core detonation. In the present study, we considered two distinct \texttt{AREPO}-generated initial conditions that were previously evolved using \texttt{AREPO}, and evolved them independently using \texttt{FLASH} at a comparable high spatial resolution  $(\sim 14 \ \mathrm{km})$. In both code frameworks, each of these models successfully obtained the core detonation of the primary. Conversely, our initial \texttt{FLASH} simulation of the \texttt{Double\_Det} model performed at a coarser spatial resolution of $26.8 \ \mathrm{km}$ failed to produce a core detonation (not shown here), despite employing identical initial conditions. 

The discrepancy in detonation outcomes between several previous models \citep{guillochon2010surface, pakmor2013helium, roy20223d}, and the models from \cite{pakmor2022fate} and present study indicates detonation outcome is likely to be sensitive not only to spatial resolution of mesh but also the method employed to construct the initial conditions. As discussed in Section \ref{sec:introduction}, the initial conditions constructed by an SPH code introduced artificially large entropy into the helium layers, puffing them up, and making them asymmetric. Also, the differential rotation of the WD during the evolution further enhanced the asymmetry of the outer layers, ablating and mixing the helium shell with carbon/oxygen core \citep{roy20223d}. Detonations in such asymmetrical and thick helium layers propagate unevenly and slowly, creating weaker shocks that converge further from the WD center. As a result, the hot spots generated by the converging shocks fail to reach sufficient conditions required for carbon ignition. For example, in the model of \cite{roy20223d}, the helium detonation phase lasts $\sim 2$ s, corresponding to a propagation speed $\lesssim 8500 \ \mathrm{km \ s^{-1}}$. In contrast, in our \texttt{Double\_Det} model, which inherits a symmetric helium structure from \texttt{AREPO}-generated initial conditions, the detonation phase lasts $\sim 1.3$ s, implying a speed of $\sim 12000 \ \mathrm{km \ s^{-1}}$. The resulting shocks from the helium detonation rapidly converge near the WD core and ignite the carbon detonation. Thus, the initial condition generation method also plays a crucial role in the detonation outcomes.

Following the explosion of primary WD in the \texttt{Double\_Det} model, the secondary is no longer gravitationally bound in the binary system and is ejected at a radial velocity on the order of a few thousand $\mathrm{km \ s^{-1}}$. This escape speed is consistent with the speed of a few D6 candidates discovered via \textit{Gaia} \citep{shen2018three, elbadryetal23}. In addition, we find that the outer layers of this surviving secondary become heavily ``polluted" by the mixing of carbon/oxygen core with nuclear ash ($\mathrm{silicon, sulfur, and \ iron}$) swept up from the primary's ejecta and surface helium burning. Similar carbon/oxygen outer shells enriched with iron and nickel have been seen in the ultraviolet spectroscopy of the hypervelocity WD J0927-6335 \citep{werner2024ultraviolet}. While the shock heating of secondary may seems plausible to explain the inflated structures seen in the hypervelocity WDs \citep{shen2018three}, there is an ongoing tension regarding the longevity of this state. Recent modeling suggests that shock-induced thermal inflation may only persist for only for a short duration compared to the inferred lifetime of the hypervelocity WDs, and that violent mergers can help resolve this tension \citep{bhat2025supernova}. 

The explosion of the primary in both the \texttt{Double\_Det} and \texttt{Quad\_Det} models will also leave distinct impressions upon the ejecta structure. The double detonation of the primary WD will produce two compositionally different shells, each imprinted with nucleosynthetic signatures set by its burning density. The surface helium detonation at densities $\sim 10^{5} \ \mathrm{g \ cm^{-3}}$ primarily yield IMEs (silicon, sulfur, calcium, and titanium), while the core detonation at higher density at around $\sim 10^{7} \mathrm{g \ cm^{-3}}$ allow for more complete burning approaching nuclear statistical equilibrium (NSE), shifting to IGE-rich (chromium, iron, and nickel) abundances. The resulting abundance stratification provides a compelling explanation for the distinct morphologies seen in the X-ray observations of the supernova remnants (SNRs). For example, the unique morphology of SNR 0509-67.5, where high-resolution mapping has recently revealed a dual shell structure dominated by ionized calcium and sulfur, are likely to be originated from the double detonation of a sub-$\mathrm{M_{Ch}}$ mass WD progenitor \citep{das2025calcium}. Additionally, during the primary WD explosion, the companion partially blocks the quasi-spherical expanding ejecta and leaves a distinct conical shadow behind the secondary WD. Over a longer evolution period $(\gtrsim 10^{2} \ \mathrm{years})$, this feature persists and evolve in a large-scale cavity deprived of IGEs in the remnant phase of the ejecta \citep{tanikawa2018three, pakmor2022fate, prust2025ejecta, ferrand2025role}. These asymmetries predicted by our models including theviewing angle dependency, may provide an important framework for explaining the structures observed in other SNRs.

Observations with current and next-generation X-ray missions will be crucial for providing high-resolution data capable of distinguishing between progenitor channels through their predicted signatures. Specifically, \textit{XRISM} mission equipped with the \texttt{RESOLVE} X-ray Calorimeter System (XCS) for soft X‑ray spectroscopy is capable of resolving Doppler‑shifted K-$\alpha$ and L-$\alpha$ lines from both IMEs and IGEs at high spectral resolution $(\lesssim 5 \ \mathrm{eV})$, which can probe asymmetric velocity structures in young SNRs \citep{ishisaki2025resolve, fujimaru2026probing}. In the near future, \textit{NewAthena} with its X-ray Integral Field Unit (X-IFU) will deliver unprecedented spatial and energy resolution. X-IFU will obtain spatially-resolved measurements of Doppler shifts, line broadening, and detailed abundance ratios that will enable full three-dimensional reconstruction of the compositional and kinematic structure of SNRs \citep{barret2016athena}.     

The quadruple detonation scenario is distinguished from the D6 scenario by the complete disruption of both the WDs in the system. The core detonation of the secondary WD in our \texttt{Quad\_Det} model increases the total ejecta mass by approximately $0.6 \ \mathrm{M_{\odot}}$, resulting in nearly $1.6 \ \mathrm{M_{\odot}}$ of burned material being released into the circumstellar environment. However, because the secondary WD has a lower mass and central density compared to the primary, its explosion is largely dominated by the production of IMEs. While this additional burning impacts the global abundance pattern--specifically enriching the ejecta with extra silicon and calcium--it has a negligible effect on the total $^{56}\mathrm{Ni}$ yield. Consequently, the explosion of the secondary WD extends the diffusion time, causing light curve to become broader in the quadrupole detonation scenario compared the the D6 scenario. In the remnant phase, the ejecta morphology also differs significantly. Unlike the D6 scenario, where the ejecta are marked by a conical cavity left by the surviving companion, the quadruple detonation does not produce such a cavity. Further, because the core detonation of the secondary WD arises after the primary core detonation and with less kinetic energy, it naturally produces a double-layered structured in the ejecta. In particular, the IMEs from the secondary WD's explosion collide with the IGE-dominant inner ejecta produced by the primary.     


Perhaps most importantly, the quadruple detonation scenario offers a potential resolution to the ongoing tension between the predicted number of hypervelocity WDs and the relatively small population actually detected in the \textit{Gaia} catalogs \citep{shen2018three, elbadryetal23, shen2025evolution}.  While the D6 scenario is considered to be a leading candidate for SNe Ia, obliterating the companion WD in quadruple detonation scenario may naturally explain the scarcity of hypervelocity WDs. The possibility that the secondary WD also explodes is supported by the ordinary stellar evolution models. These models predict that most of the sub-$\mathrm{M_{Ch}}$ carbon/oxygen WD with masses $\leq 1.0 \ \mathrm{M_{\odot}}$ emerge with a small amount of helium on their outer shells from their progenitors. Thus, a very minimal amount of accreted helium may be sufficient to ignite the helium on their surface, that can produce the double detonation \citep{shen2024almost}.



\section{Conclusion} \label{sec:conclusion}
In this study, we performed three-dimensional hydrodynamical simulations of helium-ignited binary WD merger channel. By evolving two distinct initial binary WD systems, each using two different hydrodynamical simulation frameworks--\texttt{AREPO} and \texttt{FLASH}--we have characterized the robustness of the detonation outcomes. The key findings of our study are summarized below: 
\begin{itemize}
    \item Both \texttt{AREPO} and \texttt{FLASH} simulations now consistently demonstrate both the canonical D6-scenario and the quadruple detonation scenario, depending on the initial WD and He masses. With fixed initial condition, this consistency is obtained regardless of the differences in underlying numerical methods, nuclear reaction networks, and mesh discretization strategies employed by both codes. These results strongly support the viability of both the canonical D6-scenario and the quadruple detonation scenarios from a theoretical and simulation standpoint.
    
    \item Analysis using  \texttt{FLASH} Lagrangian tracers confirm that the helium detonations of the primary WD in the \texttt{Double\_Det} and the \texttt{Quad\_Det} models are ignited primarily by the $\alpha-$capture reactions onto heavier seed nuclei $(\mathrm{^{12}C \ ( \alpha,\gamma) \ ^{16}O}, \mathrm{^{16}O \ ( \alpha,\gamma) \ ^{20}Ne})$. 

    \item For the \texttt{Double\_Det} model, the helium layer ignition time is nearly the same across both codes. However, in the \texttt{Quad\_Det} model, \texttt{FLASH} ignites approximately $\mathrm{2.8 \ seconds}$ earlier than \texttt{AREPO}. Because of this difference in the helium ignition at early stage, a phase offset persist in the subsequent evolution. This phase offset is relatively insignificant here, but may yield a greater disparity in outcomes as one approaches the threshold of core detonation. These differences between \texttt {FLASH} and \texttt{AREPO} are likely due to the the sizes of the nuclear reaction networks, which emphasizes the importance of larger nuclear networks for the accurate treatment of helium ignition.
    
    \item We also investigate the thermodynamic conditions at core shock-shock convergence. After convergence, the density and the temperature increase by more than a factor of five, and two orders of magnitude, respectively. These resulting thermodynamic conditions are sufficient to ignite carbon burning, which immediately triggers core detonation of the primary WD in both models. 


    \item We find good agreement of nuclear energy release and ejecta masses between the \texttt{AREPO} and the \texttt{FLASH} simulations, across both the models. However, we also notice some discrepancies in the abundances of heavier isotopes. These differences are more pronounced in the \texttt{Double\_det} model, specifically in $\mathrm{^{32}S}$ and $\mathrm{^{56}Ni}$ abundances. These discrepancies are likely due to the differences in the nuclear reaction networks used by both codes.   
 \end{itemize}






Nonetheless, while the models shown in our study successfully demonstrate both the D6 and quadruple detonation scenarios, the WDs in these models possess optimistically large helium shell masses $(\geq 0.005 \ \mathrm{M_{\odot}})$. Isolated stellar evolution in such binary systems points towards smaller helium masses $(\sim 10^{-3} - 10^{-5}  \mathrm{M_{\odot}})$, which are inferred from the observation of DB-class WDs \citep{kawaler1994precision, duan2021epic, corsico2021pulsating}. Among these systems, the donor WDs are likely to emerge as DA-class WDs that evolve into DB-class by the time of merger. Thus, a key constraint to push the viability of these channels under fully realistic conditions must include the need to explore binary WD systems with astrophysically-relevant thinner helium shell masses. 

\let\internallinenumbers\relax
\let\endinternallinenumbers\relax
\begin{acknowledgments}

R.T.F. acknowledges support from NSF award AST-2511516 and NASA ATP award 80NSSC22K0630. This work used the Extreme Science and Engineering Discovery Environment (XSEDE) Stampede 3 supercomputer at the University of Texas at Austin’s Texas Advanced Computing Center through allocation TG-AST100038. XSEDE is supported by National Science Foundation grant number ACI-1548562 \citep{townsetal14}. RP gratefully acknowledge the Gauss Centre for Supercomputing e.V. (www.gauss-centre.eu) for funding this project by providing computing time on the GCS Supercomputer SuperMUC-NG at Leibniz Supercomputing Centre (www.lrz.de) via the project pn76fu.
 
\end{acknowledgments}

\section{Software and third party data repository citations} \label{sec:cite}

\software{Flash 4.0 \citep{fryxell2000flash}, Arepo \citep{weinberger2020arepo}, Python programming language \citep{vanrossumdeboer91}, Yt \citep{turk_2011}, Numpy \citep{vanderwaltetal11}, Matplotlib \citep{hunter07}, Jupyter/Ipython \citep{perezgranger07}, Scipy \citep{2020SciPy-NMeth}, Paicos \citep{Berlok_Paicos_A_Python_2024}, Pynucastro \citep{pynucastro2}}


\bibliographystyle{aasjournal}
\bibliography{merged,reference_sne}

\appendix 
\section{appendix }\label{appendix}
\begin{equation}
    \frac{\rho_{2}}{\rho_{1}} = \frac{(\gamma+1)M_{1}^{2}}{(\gamma - 1)M_{1}^{2} + 2}
    \label{eq:1}
\end{equation}

Here, $\rho_{2}$ and $\rho_{1}$ denote the post-shock and pre-shock densities of the fluid, respectively. $M_{1}$ is the Mach number of the shock, which depends on the local sound speed in the medium, and $\gamma$ is the adiabatic index of the gas. In the presence of very strong shock $(M_{1} \gg 1)$, the density ratio is approximate to $\frac{\gamma + 1}{\gamma - 1}$, which corresponds to the factor of four for non-relativistic degenerate gas $(\gamma = \frac{5}{3})$. Considering a similar ideal strong shock case for our simulation, we estimated $\frac{\rho_{2}}{\rho_{1}} \approx 5.65 $ for the $\gamma = 1.43$ of that tracer, which is less than the ratios we get from our analysis. To interpret this excess, we next consider the case of a strong shock incident over a symmetry plane (or equivalently, two identical shocks converging symmetrically). In this framework, the incident shock is effectively reflected at the boundary (or meets its counterpart), producing a reflected shock that propagates back into the already shocked region at Mach number $M_{R}$. For a strong incident shock, the maximum Mach number of the reflected shock is given by Equation \eqref{eq:2}, which we use to estimate the additional compression at the point of shock–shock convergence.


\begin{equation}
    M^{max}_{R} = \sqrt{\frac{2\gamma}{\gamma - 1}} 
    \label{eq:2}
\end{equation}

We can again use Rankine-Hugoniot relation for the reflected shock as shown in equation \eqref{eq:3}.

\begin{equation}
    \frac{\rho_{3}}{\rho_{2}} = \frac{(\gamma+1)M_{R}^{2}}{(\gamma - 1)M_{R}^{2} + 2}
    \label{eq:3}
\end{equation}

Here $\rho_{2}$ and $\rho_{3}$ denote the post-incident shock and post-reflection shock densities. We now substitute for $M_{R} = M_{R}^{max}$ and simplify.

\begin{equation}
    \frac{\rho_{3}}{\rho_{2}} = \frac{(\gamma+1)(M^{max}_{R})^{2}}{(\gamma - 1)(M^{max}_{R})^{2} + 2} 
    = \frac{(\gamma+1)\left(\frac{2\gamma}{\gamma - 1}\right)}{(\gamma - 1)\left(\frac{2\gamma}{\gamma - 1}\right) + 2} 
    = \frac{2\gamma (\gamma+1)}{2(\gamma+1)(\gamma - 1)}
    = \frac{\gamma}{\gamma-1}
    \label{eq:4}
\end{equation}

Combining equations \eqref{eq:4} and \eqref{eq:1} for strong incident shock $(M_{1} \gg 1)$, yields the maximum theoretical density amplification due to shock–shock convergence.

\begin{equation}
   \frac{\rho_{3}}{\rho_{1}} = \left(\frac{\gamma + 1}{\gamma - 1}\right) \left( \frac{\gamma}{\gamma - 1}\right) = \frac{\gamma(\gamma+1)}{(\gamma - 1)^{2}} 
   \label{eq:5}
\end{equation}

Using the adiabatic index $\gamma = 5/3$, Equation \eqref{eq:5} predicts a maximum density amplification of $\rho_{3}/\rho_{1} \approx 10$ for two converging shocks, while for the $\gamma = 1.43$ of the tracer in this simulation, the equation \eqref{eq:5} gives $\rho_{3}/\rho_{1} \approx 18.8$. 



\end{document}